\documentstyle[12pt,aasms4]{article}
\begin{document}
%
\typeout{TCILATEX Macros for Scientific Word 1.1 <09 Jun 93>.}
%
%
%
\def\BF#1{{\bf {#1}}}%
\def\NEG#1{{\rlap/#1}}%
%
%
\makeatletter
%
%
%
\let\DOTSI\relax
\def\RIfM@{\relax\ifmmode}%
\def\FN@{\futurelet\next}%
\newcount\intno@
\def\iint{\DOTSI\intno@\tw@\FN@\ints@}%
\def\iiint{\DOTSI\intno@\thr@@\FN@\ints@}%
\def\iiiint{\DOTSI\intno@4 \FN@\ints@}%
\def\idotsint{\DOTSI\intno@\z@\FN@\ints@}%
\def\ints@{\findlimits@\ints@@}%
\newif\iflimtoken@
\newif\iflimits@
\def\findlimits@{\limtoken@true\ifx\next\limits\limits@true
 \else\ifx\next\nolimits\limits@false\else
 \limtoken@false\ifx\ilimits@\nolimits\limits@false\else
 \ifinner\limits@false\else\limits@true\fi\fi\fi\fi}%
\def\multint@{\int\ifnum\intno@=\z@\intdots@                                
 \else\intkern@\fi                                                          
 \ifnum\intno@>\tw@\int\intkern@\fi                                         
 \ifnum\intno@>\thr@@\int\intkern@\fi                                       
 \int}
\def\multintlimits@{\intop\ifnum\intno@=\z@\intdots@\else\intkern@\fi
 \ifnum\intno@>\tw@\intop\intkern@\fi
 \ifnum\intno@>\thr@@\intop\intkern@\fi\intop}%
\def\intic@{\mathchoice{\hskip.5em}{\hskip.4em}{\hskip.4em}{\hskip.4em}}%
\def\negintic@{\mathchoice
 {\hskip-.5em}{\hskip-.4em}{\hskip-.4em}{\hskip-.4em}}%
\def\ints@@{\iflimtoken@                                                    
 \def\ints@@@{\iflimits@\negintic@\mathop{\intic@\multintlimits@}\limits    
  \else\multint@\nolimits\fi                                                
  \eat@}
 \else                                                                      
 \def\ints@@@{\iflimits@\negintic@
  \mathop{\intic@\multintlimits@}\limits\else
  \multint@\nolimits\fi}\fi\ints@@@}%
\def\intkern@{\mathchoice{\!\!\!}{\!\!}{\!\!}{\!\!}}%
\def\plaincdots@{\mathinner{\cdotp\cdotp\cdotp}}%
\def\intdots@{\mathchoice{\plaincdots@}%
 {{\cdotp}\mkern1.5mu{\cdotp}\mkern1.5mu{\cdotp}}%
 {{\cdotp}\mkern1mu{\cdotp}\mkern1mu{\cdotp}}%
 {{\cdotp}\mkern1mu{\cdotp}\mkern1mu{\cdotp}}}%
%
%
%
\def\rmfam{\z@}%
\newif\iffirstchoice@
\firstchoice@true
\def\textfonti{\the\textfont\@ne}%
\def\textfontii{\the\textfont\tw@}%
\def\text{\RIfM@\expandafter\text@\else\expandafter\text@@\fi}%
\def\text@@#1{\leavevmode\hbox{#1}}%
\def\text@#1{\mathchoice
 {\hbox{\everymath{\displaystyle}\def\textfonti{\the\textfont\@ne}%
  \def\textfontii{\the\textfont\tw@}\textdef@@ T#1}}%
 {\hbox{\firstchoice@false
  \everymath{\textstyle}\def\textfonti{\the\textfont\@ne}%
  \def\textfontii{\the\textfont\tw@}\textdef@@ T#1}}%
 {\hbox{\firstchoice@false
  \everymath{\scriptstyle}\def\textfonti{\the\scriptfont\@ne}%
  \def\textfontii{\the\scriptfont\tw@}\textdef@@ S\rm#1}}%
 {\hbox{\firstchoice@false
  \everymath{\scriptscriptstyle}\def\textfonti
  {\the\scriptscriptfont\@ne}%
  \def\textfontii{\the\scriptscriptfont\tw@}\textdef@@ s\rm#1}}}%
\def\textdef@@#1{\textdef@#1\rm\textdef@#1\bf\textdef@#1\sl\textdef@#1\it}%
\def\DN@{\def\next@}%
\def\eat@#1{}%
\def\textdef@#1#2{%
 \DN@{\csname\expandafter\eat@\string#2fam\endcsname}%
 \if S#1\edef#2{\the\scriptfont\next@\relax}%
 \else\if s#1\edef#2{\the\scriptscriptfont\next@\relax}%
 \else\edef#2{\the\textfont\next@\relax}\fi\fi}%
%
%
%
\def\Let@{\relax\iffalse{\fi\let\\=\cr\iffalse}\fi}%
\def\vspace@{\def\vspace##1{\crcr\noalign{\vskip##1\relax}}}%
\def\multilimits@{\bgroup\vspace@\Let@
 \baselineskip\fontdimen10 \scriptfont\tw@
 \advance\baselineskip\fontdimen12 \scriptfont\tw@
 \lineskip\thr@@\fontdimen8 \scriptfont\thr@@
 \lineskiplimit\lineskip
 \vbox\bgroup\ialign\bgroup\hfil$\m@th\scriptstyle{##}$\hfil\crcr}%
\def\Sb{_\multilimits@}%
\def\endSb{\crcr\egroup\egroup\egroup}%
\def\Sp{^\multilimits@}%
\let\endSp\endSb
%
%
%
\newdimen\ex@
\ex@.2326ex
\def\rightarrowfill@#1{$#1\m@th\mathord-\mkern-6mu\cleaders
 \hbox{$#1\mkern-2mu\mathord-\mkern-2mu$}\hfill
 \mkern-6mu\mathord\rightarrow$}%
\def\leftarrowfill@#1{$#1\m@th\mathord\leftarrow\mkern-6mu\cleaders
 \hbox{$#1\mkern-2mu\mathord-\mkern-2mu$}\hfill\mkern-6mu\mathord-$}%
\def\leftrightarrowfill@#1{$#1\m@th\mathord\leftarrow\mkern-6mu\cleaders
 \hbox{$#1\mkern-2mu\mathord-\mkern-2mu$}\hfill
 \mkern-6mu\mathord\rightarrow$}%
\def\overrightarrow{\mathpalette\overrightarrow@}%
\def\overrightarrow@#1#2{\vbox{\ialign{##\crcr\rightarrowfill@#1\crcr
 \noalign{\kern-\ex@\nointerlineskip}$\m@th\hfil#1#2\hfil$\crcr}}}%
\let\overarrow\overrightarrow
\def\overleftarrow{\mathpalette\overleftarrow@}%
\def\overleftarrow@#1#2{\vbox{\ialign{##\crcr\leftarrowfill@#1\crcr
 \noalign{\kern-\ex@\nointerlineskip}$\m@th\hfil#1#2\hfil$\crcr}}}%
\def\overleftrightarrow{\mathpalette\overleftrightarrow@}%
\def\overleftrightarrow@#1#2{\vbox{\ialign{##\crcr\leftrightarrowfill@#1\crcr
 \noalign{\kern-\ex@\nointerlineskip}$\m@th\hfil#1#2\hfil$\crcr}}}%
\def\underrightarrow{\mathpalette\underrightarrow@}%
\def\underrightarrow@#1#2{\vtop{\ialign{##\crcr$\m@th\hfil#1#2\hfil$\crcr
 \noalign{\nointerlineskip}\rightarrowfill@#1\crcr}}}%
\let\underarrow\underrightarrow
\def\underleftarrow{\mathpalette\underleftarrow@}%
\def\underleftarrow@#1#2{\vtop{\ialign{##\crcr$\m@th\hfil#1#2\hfil$\crcr
 \noalign{\nointerlineskip}\leftarrowfill@#1\crcr}}}%
\def\underleftrightarrow{\mathpalette\underleftrightarrow@}%
\def\underleftrightarrow@#1#2{\vtop{\ialign{##\crcr$\m@th\hfil#1#2\hfil$\crcr
 \noalign{\nointerlineskip}\leftrightarrowfill@#1\crcr}}}%
%
%
\def\tfrac#1#2{{\textstyle {#1 \over #2}}}%
\def\dfrac#1#2{{\displaystyle {#1 \over #2}}}%
\def\binom#1#2{{#1 \choose #2}}%
\def\tbinom#1#2{{\textstyle {#1 \choose #2}}}%
\def\dbinom#1#2{{\displaystyle {#1 \choose #2}}}%
\def\QATOP#1#2{{#1 \atop #2}}%
\def\QTATOP#1#2{{\textstyle {#1 \atop #2}}}%
\def\QDATOP#1#2{{\displaystyle {#1 \atop #2}}}%
\def\QABOVE#1#2#3{{#2 \above#1 #3}}%
\def\QTABOVE#1#2#3{{\textstyle {#2 \above#1 #3}}}%
\def\QDABOVE#1#2#3{{\displaystyle {#2 \above#1 #3}}}%
\def\QOVERD#1#2#3#4{{#3 \overwithdelims#1#2 #4}}%
\def\QTOVERD#1#2#3#4{{\textstyle {#3 \overwithdelims#1#2 #4}}}%
\def\QDOVERD#1#2#3#4{{\displaystyle {#3 \overwithdelims#1#2 #4}}}%
\def\QATOPD#1#2#3#4{{#3 \atopwithdelims#1#2 #4}}%
\def\QTATOPD#1#2#3#4{{\textstyle {#3 \atopwithdelims#1#2 #4}}}%
\def\QDATOPD#1#2#3#4{{\displaystyle {#3 \atopwithdelims#1#2 #4}}}%
\def\QABOVED#1#2#3#4#5{{#4 \abovewithdelims#1#2#3 #5}}%
\def\QTABOVED#1#2#3#4#5{{\textstyle {#4 \abovewithdelims#1#2#3 #5}}}%
\def\QDABOVED#1#2#3#4#5{{\displaystyle {#4 \abovewithdelims#1#2#3 #5}}}%
%
%
\def\tint{\textstyle \int}%
\def\tiint{\mathop{\textstyle \iint }}%
\def\tiiint{\mathop{\textstyle \iiint }}%
\def\tiiiint{\mathop{\textstyle \iiiint }}%
\def\tidotsint{\mathop{\textstyle \idotsint }}%
\def\toint{\textstyle \oint}%
\def\tsum{\mathop{\textstyle \sum }}%
\def\tprod{\mathop{\textstyle \prod }}%
\def\tbigcap{\mathop{\textstyle \bigcap }}%
\def\tbigwedge{\mathop{\textstyle \bigwedge }}%
\def\tbigoplus{\mathop{\textstyle \bigoplus }}%
\def\tbigodot{\mathop{\textstyle \bigodot }}%
\def\tbigsqcup{\mathop{\textstyle \bigsqcup }}%
\def\tcoprod{\mathop{\textstyle \coprod }}%
\def\tbigcup{\mathop{\textstyle \bigcup }}%
\def\tbigvee{\mathop{\textstyle \bigvee }}%
\def\tbigotimes{\mathop{\textstyle \bigotimes }}%
\def\tbiguplus{\mathop{\textstyle \biguplus }}%
%
%
%
\def\dint{\displaystyle \int }%
\def\diint{\mathop{\displaystyle \iint }}%
\def\diiint{\mathop{\displaystyle \iiint }}%
\def\diiiint{\mathop{\displaystyle \iiiint }}%
\def\didotsint{\mathop{\displaystyle \idotsint }}%
\def\doint{\displaystyle \oint }%
\def\dsum{\mathop{\displaystyle \sum }}%
\def\dprod{\mathop{\displaystyle \prod }}%
\def\dbigcap{\mathop{\displaystyle \bigcap }}%
\def\dbigwedge{\mathop{\displaystyle \bigwedge }}%
\def\dbigoplus{\mathop{\displaystyle \bigoplus }}%
\def\dbigodot{\mathop{\displaystyle \bigodot }}%
\def\dbigsqcup{\mathop{\displaystyle \bigsqcup }}%
\def\dcoprod{\mathop{\displaystyle \coprod }}%
\def\dbigcup{\mathop{\displaystyle \bigcup }}%
\def\dbigvee{\mathop{\displaystyle \bigvee }}%
\def\dbigotimes{\mathop{\displaystyle \bigotimes }}%
\def\dbiguplus{\mathop{\displaystyle \biguplus }}%
%
\def\stackunder#1#2{\mathrel{\mathop{#2}\limits_{#1}}}%
%
%
%
\def\FILENAME#1{#1}%
\newcount\GRAPHICSTYPE
\GRAPHICSTYPE=\z@
\def\GRAPHICSPS#1{%
 \ifcase\GRAPHICSTYPE
  ps: #1%
 \or
  language "PS", include "#1"%
 \or
  #1%
 \fi
}%
\def\GRAPHICSHP#1{include #1}%
%
\def\graffile#1#2#3#4{%
 \ifnum\GRAPHICSTYPE=\tw@
  \@ifundefined{psfig}{\input psfig.tex}{}%
  \psfig{file=#1, height=#3, width=#2}%
 \else
  \leavevmode\raise -#4 \hbox{%
   \raise #3 \hbox{\rule{0.003in}{0.003in}\special{#1}}%
   }%
  {\raise -#4 \hbox to #2 {\vrule height#3 width\z@ depth\z@\hfil}}%
 \fi
}%
%
\def\draftbox#1#2#3#4{%
 \leavevmode\raise -#4 \hbox{%
  \frame{\rlap{\protect\tiny #1}\hbox to #2%
   {\vrule height#3 width\z@ depth\z@\hfil}%
  }%
 }%
}%
\newcount\draft
\draft=\z@
\def\GRAPHIC#1#2#3#4#5{%
 \ifnum\draft=\@ne\draftbox{#2}{#3}{#4}{#5}%
  \else\graffile{#1}{#3}{#4}{#5}%
  \fi
 }%
\def\addtoLaTeXparams#1{\edef\LaTeXparams{\LaTeXparams #1}}%
\def\doFRAMEparams#1{\readFRAMEparams#1\end}%
\def\readFRAMEparams#1{%
 \ifx#1\end%
  \let\next=\relax
  \else
  \ifx#1i\dispkind=\z@\fi
  \ifx#1d\dispkind=\@ne\fi
  \ifx#1f\dispkind=\tw@\fi
  \ifx#1t\addtoLaTeXparams{t}\fi
  \ifx#1b\addtoLaTeXparams{b}\fi
  \ifx#1p\addtoLaTeXparams{p}\fi
  \ifx#1h\addtoLaTeXparams{h}\fi
  \let\next=\readFRAMEparams
  \fi
 \next
 }%
%
\def\IFRAME#1#2#3#4#5{\GRAPHIC{#5}{#4}{#1}{#2}{#3}}%
%
\def\DFRAME#1#2#3#4{%
 \begin{center}\GRAPHIC{#4}{#3}{#1}{#2}{\z@}\end{center}%
 }%
%
\def\FFRAME#1#2#3#4#5#6#7{%
 \begin{figure}[#1]%
  \begin{center}\GRAPHIC{#7}{#6}{#2}{#3}{\z@}\end{center}%
  \caption{\label{#5}#4}%
  \end{figure}%
 }%
%
%
%
%
%
\newcount\dispkind%
\def\FRAME#1#2#3#4#5#6#7#8{%
 \def\LaTeXparams{}%
 \dispkind=\z@
 \def\LaTeXparams{}%
 \doFRAMEparams{#1}%
 \ifnum\dispkind=\z@\IFRAME{#2}{#3}{#4}{#7}{#8}\else
  \ifnum\dispkind=\@ne\DFRAME{#2}{#3}{#7}{#8}\else
   \ifnum\dispkind=\tw@
    \edef\@tempa{\noexpand\FFRAME{\LaTeXparams}}%
    \@tempa{#2}{#3}{#5}{#6}{#7}{#8}%
    \fi
   \fi
  \fi
 }%
%
\def\func#1{\mathop{\rm #1}}%
\def\limfunc#1{\mathop{\rm #1}}%
%
\long\def\QQQ#1#2{\long\expandafter\def\csname#1\endcsname{#2}}%
\def\QTP#1{}%
\def\QWE{}%
\long\def\QQA#1#2{}%
\def\QTR#1#2{{\csname#1\endcsname #2}}
\long\def\TeXButton#1#2{#2}%
\def\EXPAND#1[#2]#3{}%
\def\NOEXPAND#1[#2]#3{}%
\def\PROTECTED{}%
\def\LaTeXparent#1{}%
\def\QTagDef#1#2#3{}%
%
\def\QQfnmark#1{\footnotemark}
\def\QQfntext#1#2{\addtocounter{footnote}{#1}\footnotetext{#2}}
%
\def\MAKEINDEX{\makeatletter\input gnuindex.sty\makeatother\makeindex}%
\@ifundefined{INDEX}{\def\INDEX#1#2{}{}}{}%
\@ifundefined{SUBINDEX}{\def\SUBINDEX#1#2#3{}{}{}}{}%
\def\initial#1{\bigbreak{\raggedright\large\bf #1}\kern 2\p@\penalty3000}%
\def\entry#1#2{\item {#1}, #2}%
\def\primary#1{\item {#1}}%
\def\secondary#1#2{\subitem {#1}, #2}%
%
\@ifundefined{abstract}{%
 \def\abstract{%
  \if@twocolumn
   \section*{Abstract (Not appropriate in this style!)}%
   \else \small 
   \begin{center}{\bf Abstract\vspace{-.5em}\vspace{\z@}}\end{center}%
   \quotation 
   \fi
  }%
 }{%
 }%
\@ifundefined{endabstract}{\def\endabstract
  {\if@twocolumn\else\endquotation\fi}}{}%
\@ifundefined{maketitle}{\def\maketitle#1{}}{}%
\@ifundefined{affiliation}{\def\affiliation#1{}}{}%
\@ifundefined{proof}{\def\proof{\paragraph{Proof. }}}{}%
\@ifundefined{endproof}{\def\endproof{\mbox{\ $\Box$}}}{}%
\@ifundefined{newfield}{\def\newfield#1#2{}}{}%
\@ifundefined{chapter}{\def\chapter#1{\par(Chapter head:)#1\par }%
 \newcount\c@chapter}{}%
\@ifundefined{part}{\def\part#1{\par(Part head:)#1\par }}{}%
\@ifundefined{section}{\def\section#1{\par(Section head:)#1\par }}{}%
\@ifundefined{subsection}{\def\subsection#1%
 {\par(Subsection head:)#1\par }}{}%
\@ifundefined{subsubsection}{\def\subsubsection#1%
 {\par(Subsubsection head:)#1\par }}{}%
\@ifundefined{paragraph}{\def\paragraph#1%
 {\par(Subsubsubsection head:)#1\par }}{}%
\@ifundefined{subparagraph}{\def\subparagraph#1%
 {\par(Subsubsubsubsection head:)#1\par }}{}%
%
\@ifundefined{therefore}{\def\therefore{}}{}%
\@ifundefined{backepsilon}{\def\backepsilon{}}{}%
\@ifundefined{yen}{\def\yen{\hbox{\rm\rlap=Y}}}{}%
\@ifundefined{registered}{\def\registered{\relax\ifmmode{}\r@gistered
                                                \else$\m@th\r@gistered$\fi}%
 \def\r@gistered{^{\ooalign
  {\hfil\raise.07ex\hbox{$\scriptstyle\rm\text{R}$}\hfil\crcr
  \mathhexbox20D}}}}{}%
\@ifundefined{Eth}{\def\Eth{}}{}%
\@ifundefined{eth}{\def\eth{}}{}%
\@ifundefined{Thorn}{\def\Thorn{}}{}%
\@ifundefined{thorn}{\def\thorn{}}{}%
\def\TEXTsymbol#1{\mbox{$#1$}}%
\@ifundefined{degree}{\def\degree{{}^{\circ}}}{}%
%
\def\BibTeX{{\rm B\kern-.05em{\sc i\kern-.025em b}\kern-.08em
    T\kern-.1667em\lower.7ex\hbox{E}\kern-.125emX}}%
%
\newdimen\theight
\def\Column{%
 \vadjust{\setbox\z@=\hbox{\scriptsize\quad\quad tcol}%
  \theight=\ht\z@\advance\theight by \dp\z@\advance\theight by \lineskip
  \kern -\theight \vbox to \theight{%
   \rightline{\rlap{\box\z@}}%
   \vss
   }%
  }%
 }%
\def\qed{%
 \ifhmode\unskip\nobreak\fi\ifmmode\ifinner\else\hskip5\p@\fi\fi
 \hbox{\hskip5\p@\vrule width4\p@ height6\p@ depth1.5\p@\hskip\p@}%
 }%
\def\cents{\hbox{\rm\rlap/c}}%
\def\miss{\hbox{\vrule height2\p@ width 2\p@ depth\z@}}%
\def\vvert{\Vert}
\def\tcol#1{{\baselineskip=6\p@ \vcenter{#1}} \Column}  %
\def\dB{\hbox{{}}}
\def\mB#1{\hbox{$#1$}}
\def\nB#1{\hbox{#1}}
%
\def\note{$^{\dag}}%
\makeatother
%
%
\newcommand{\Grad}[1]{\mbox{$\nabla #1$}}
\newcommand{\Div} [1]{\mbox{$\nabla\hspace{-3pt}\cdot\hspace{-2pt}#1$}}
\newcommand{\Curl}[1]{\mbox{$\nabla\hspace{-3pt}\times\hspace{-2pt}#1$}}
\newcommand{\Lap} [1]{\mbox{$\nabla^{2} #1$}}
\def\gapprox{\mathrel{\vcenter{\offinterlineskip \hbox{$>$}
    \kern 0.3ex \hbox{$\sim$}}}}
\setlength{\baselineskip}{12pt}

\title{MHD Models of Axisymmetric Protostellar Jets}

\author{James M. Stone}
\affil{Department of Astronomy, University of Maryland, College Park,
MD 20742 \\ jstone@astro.umd.edu}

\author{Philip E. Hardee}
\affil{Department of Physics \& Astronomy, The University of Alabama,
Tuscaloosa, AL 35487 \\ hardee@athena.astr.ua.edu}

\begin{abstract}

We present the results of a series of axisymmetric time-dependent
magnetohydrodynamic (MHD) simulations of the propagation of cooling,
overdense jets.  Our numerical models are motivated by the properties
of outflows associated with young stellar objects.  A variety of
initial field strengths and configurations are explored for both steady
and time-variable (pulsed) jets.  For the parameters of protostellar
jets adopted here, even apparently weak magnetic fields with strengths
$B \gapprox 60 \mu$G in the pre-shocked jet beam can have a significant
effect on the dynamics, for example by altering the density, width, and
fragmentation of thin shells formed by cooling gas.  Strong toroidal
fields ($\geq 100\mu$G) with a radial profile that peaks near the
surface of the jet result in the accumulation of dense shocked gas in a
``nose cone'' at the head of jet.  We suggest that this structure is
unstable in three-dimensions.  A linear analysis predicts that
axisymmetric pinch modes of the MHD Kelvin-Helmholtz instability should
grow only slowly for the highly supermagnetosonic jets studied here; we
find no evidence for them in our simulations.  Some of our models
appear unstable to current-driven pinch modes, however the resulting
pressure and density variations induced in the jet beam are not large,
making this mechanism an unlikely source of emission knots in the jet
beam.  In the case of pulsed jets, radial hoop stresses confine shocked
jet material in the pulses to the axis, resulting in a higher density
in the pulses in comparison to purely hydrodynamic models. In addition,
if the magnetic field strength varies with radius, significant radial
structure is produced in the pulses (the density is strongly axially
peaked, for example) even if the density and velocity in the jet follow
a constant ``top-hat" profile initially.

\end{abstract}


\section{Introduction}

The most promising mechanism for the production of supersonic, highly
collimated jets from low mass young stellar objects is by magnetic forces
associated either directly with an accretion disk (K\"{o}nigl \&
Pudritz 1999), or with the interaction between an accretion disk and a
magnetized central star (Shu et al.\ 1999).  Magnetic fields are also
thought to contribute to the collimation of the jets on larger scales,
(although only slowly so that the observed jet may only be a part of a
much broader wind, e.g., Ostriker 1997).  Thus, unless the outflowing
material is highly resistive (which seems unlikely), protostellar jets
should contain a dynamically important magnetic field which may affect
both the propagation and stability of the outflow.  Observation of the
magnetic field strength associated with protostellar jets is
difficult.  However, by fitting one-dimensional radiative shock models
to the observed line ratios in the bow shock of HH47, Morse et
al.\ (1993) inferred an upper limit to the magnetic field in the
ambient medium upstream of the jet of $\sim 30 \mu$G, a value which
they argued was too small to affect the dynamics except by increasing
the cooling lengths behind radiative shocks.  More recently, Ray et al.\ (1997)
have reported direct evidence for strong fields in the outflow
associated with the source T~Tau~S through the detection of opposite
circular polarization in the two spatially resolved outflow lobes.
These authors infer very high field strengths within the lobes; several
Gauss at a distance of few tens of AU from the central star.

If protostellar jets contain strong magnetic fields as expected from
theory, there may be a signature of such fields in their dynamics.
While hydrodynamical studies of
cooling, dense protostellar jets are widely available in the
literature (e.g., see Raga 1995; Cabrit 1997 for reviews), MHD models
are less well explored.  MHD studies of extragalactic (i.e.,
{\em underdense} and {\em non-radiative}) jets have been
reported by Clarke, Norman, \& Burns (1986), and Lind et al.\ (1989) for 
the case of toroidal fields, and by K\"{o}ssl et al.\ (1990a; 1990b) for the
case of toroidal and axial fields (see also Clarke 1996 for a recent
review).  For strong fields, characterized by
a small value of $\beta \equiv 8 \pi P/B^{2}$ where B is the field
strength and P the thermal pressure (although note such fields may
still have an energy density small compared to the kinetic energy density
in the flow), it is found that the cocoon formed
by lateral expansion of hot, shocked gas from the head of the jet is
strongly inhibited.  Instead, a magnetically confined ``nose cone" of
shock processed jet material is formed between the bow shock and Mach
disk.  In addition, the stability properties of the jet beam itself are
strongly altered by the presence of a magnetic field.  MHD studies of
the propagation of {\em overdense}, non-radiative jets into a uniform ambient
medium which contains a helical magnetic field everywhere have been
reported by Todo et al.\ (1992).  They describe results for a number of
field strengths and approximate the effects of radiative cooling with
an equation of state with $\gamma=1.2$.  The formation of a nose cone,
as well as suppression of the cocoon, is evident for strong fields.  In
subsequent three-dimensional simulations (Todo et al.\ 1993), it was
found that a magnetic field strength of $70\mu$G was sufficient to
produce nonaxisymmetric kink instabilities in the jet beam which the
authors suggest may be relevant to bends observed in the jets from some
protostellar systems.  This field strength is approaching the value
inferred by Morse et al. (1993).

More recently, several authors have reported results from
MHD studies of dense jets in which optically thin radiative cooling is
directly coupled to the dynamics.  Frank et al.\ (1998, hereafter FRJN-C) report
axisymmetric studies which confirm the formation of nose cones, and
(for the particular form of the initial field distribution adopted in
the jet) the presence of pinch modes driven by magnetic tension.
Cerqueira et al.\ (1997, hereafter CGH; see also de Gouveia Dal Pino
\& Cerqueira 1996) have used an SPH code which has been extended with
an algorithm designed to represent magnetic forces to study the
propagation of dense, cooling jets in three-dimensions.  They find that
the fragmentation of the dense shell formed by cooling at the head of
the jet is strongly affected by magnetic forces, and they use these
results to conclude that the clumpy structure observed in jets provides
evidence that near the head of the jet
the field must be primarily axial rather than helical.
In their latest simulations (Cerqueira \& de Gouveia Dal Pino 1999,
hereafter CG), these authors also report evidence for both 
modes of the MHD Kelvin-Helmholtz (KH) and a current-driven pinch
instability in models of protostellar jets.
Finally, Frank et al. (1999a; see also Lery \& Frank 1999) 
have studied the propagation of MHD jets launched by Keplerian 
accretion disks using analytic models for magneto-centrifugal
outflows to describe the jet structure at the nozzle.
A variety of complex behavior is seen in this case, some
of which results from MHD effects and some which results from the
non-uniform radial structure of the jet beam.

Given the apparent sensitivity of jet dynamics to the assumed initial
orientation and strength of the magnetic field within the jet beam
suggested by published numerical results, it is clearly useful to
survey different initial field profiles.  In this paper, we present the
results from a large number of two-dimensional, axisymmetric MHD
simulations of dense, cooling jets.  To emphasize the effects produced
solely by the magnetic field, we study jets with a uniform (``top-hat")
radial profile propagating into
an initially uniform ambient medium.  Since the theory of the production
of such jets suggests a primarily toroidal field within the outflow
far from the source, we focus on
toroidal fields with different strengths and radial profiles.  In most
of our models, the ambient medium into which the jet propagates is
unmagnetized, although in some cases we add a poloidal field which
threads both the jet and ambient gas initially. 
Given the importance of nonaxisymmetric instabilities to the dynamics of
magnetized jets (Todo et al.\ 1993), fully three-dimensional MHD
simulations of cooling jets are important (CGH; CG),
however in order to cover parameter space efficiently we confine the
models described in this study to axisymmetry.  Moreover, for the
highly supermagnetosonic jets studied here, non-axisymmetric modes
of, e.g., the K-H instability grow slowly and will not affect the
jet in regions close to the nozzle.  Of course, precession of the jet
beam, a clumpy ambient medium, or a much lower magnetosonic Mach number
could introduce three-dimensional effects much earlier.
Our results extend
previous studies in that we evolve the jet for much longer (using a grid
roughly three times larger than used by
FRJN-C or CG), and we focus our attention on strongly magnetized
jet beams propagating into an unmagnetized ambient medium.
Where there is overlap, we compare and contrast
our results to previous studies throughout this paper.

The paper is organized as follows.  We describe our numerical
techniques in \S2.  As an aid to the interpretation of our results, we
review the relevant stability properties of MHD jets in \S3.  Our
results for steady are presented in \S4, and for pulsed jets in \S5.
In \S6 we summarize and conclude.

\section{Method}

\subsection{MHD Algorithms}

The dynamical evolution of a non-relativistic magnetized jet is given by
solutions to the equations of MHD:
\begin{equation}
 \frac{D \rho}{D t} + \rho \nabla \cdot {\bf v} = 0,
\end{equation}
\begin{equation}
 \rho \frac{D{\bf v}}{Dt} = - \nabla p + (\Curl{\bf B} \times {\bf B})/ 4 \pi
\end{equation}
\begin{equation}
 \rho \frac{D}{Dt} \frac{e}{\rho} =  - p \nabla \cdot {\bf v} - 
  \Lambda + H,
\end{equation}
\begin{equation}
 \frac{\partial {\bf B}}{\partial t}=\nabla \times
({\bf v} \times {\bf B}).
\end{equation}
In these equations, $D/Dt = \partial / \partial t +
{\bf v} \cdot \nabla$ is the convective derivative, $\Lambda$ and H
represent cooling and heating terms, respectively, and the other
symbols have their usual meanings.  We assume an ideal gas law for the
equation of state; that is, $p = (\gamma - 1)e$, where $\gamma$ = 5/3,
and $e$ is the internal energy density.  We adopt the assumptions of
ideal MHD in writing equations (1) -- (4), in particular we assume the
magnetic field is perfectly coupled to the fluid.  While temperatures
in the jet beam and in the strong shocks at the head of the jet are
probably high enough (and therefore the gas is ionized enough) to keep
the field well coupled on sub-parsec scales (Frank et al 1999b),
it is possible that in slower shocks near the
wings of the outflow, non-ideal MHD effects may become important,
especially in dense ambient gas.  However, the study of non-ideal MHD
effects (for example, the structure of ``C-type'' shocks, Draine
\& McKee 1993) in the
outflow is beyond the scope of this paper.

The second and third terms on the right-hand-side of equation (3)
represent energy loss via optically thin cooling and heating
respectively.  An accurate representation of these terms is one of the
most important, and yet challenging, aspects of dynamical studies of
cooling jets.  Previous work has shown significant differences in the
evolution of cooling jets using coronal versus non-equilibrium cooling
rates (Stone \& Norman 1993).  Recently, Raga et al.\ (1997) have
developed an accurate and efficient formulation to approximate
non-equilibrium cooling rates in interstellar gas.   However, given that the
focus of this paper is the effect of magnetic fields on the propagation
of cooling jets, we have chosen to adopt only a simple formulation for the
optically thin cooling rate:  the coronal cooling curve for
interstellar gas calculated by Dalgarno \& McCray (1972).  We allow gas
to cool to a temperature of 100~K, below this we set the cooling rate to
zero.

We use the ZEUS code (Stone \& Norman 1992a; 1992b) to solve the MHD
equations in cylindrical coordinates assuming axisymmetry.  The stiff
cooling and heating terms are differenced implicitly, resulting in a
nonlinear equation at every grid point which is solved using a
Newton-Raphson iteration scheme.  This step is operator split from the
rest of the MHD equations.  No other additions or modifications to
the ZEUS algorithm were required.

\subsection{Initial and Boundary Conditions}

The axisymmetric simulations are performed on a grid of size $0\leq
r\leq 20R_j$ in the transverse direction, and $0\leq z\leq 100R_j$ in
the axial direction, where $R_j$ is the initial jet radius. Except
where indicated we use 400 uniform zones in the transverse direction
and 2000 uniform zones in the axial direction, so that 20 zones span
the initial jet radius.  Reflecting boundary conditions are used along
the inner $r$ boundary, while outflow boundary conditions are used
along the outer $r$, outer $z$, and inner $z$ boundary for $r>R_j$. For
$r\leq R_j$ along the inner $z$ boundary, inflow boundary conditions
are used with the variables held fixed at the values appropriate to the
initial equilibrium structure of the jet.

Initially the jet is assumed to be perfectly collimated. Motivated by
observations of various protostellar jets (e.g., Ray 1996), we adopt
the following values as characteristic of observed systems:  an initial
jet radius $R_j = 2.5\times 10^{15}$~cm, a uniform density
$n_{j}=1000$~cm$^{-3}$, and a uniform jet velocity
$v_{j}=332$~km~s$^{-1}$.  The gas pressure at the surface of the jet is
taken to be $p_0=1.38\times 10^{-10}$ dyne cm$^{-2}$, corresponding to
a temperature of $T_{j}=10^{3}$~K and sound speed
$a_{j}=5.25$~km~s$^{-1}$.  As described below, in order to ensure
magnetostatic equilibrium in the radial direction, in some simulations
we adopt a radially varying pressure profile.  The ambient medium is
stationary, with uniform density $n_{a}=100$~cm$^{-3}$ (so that the jet
is overdense, i.e.  $\eta \equiv n_{j}/n_{a}=10$), and temperature
$T_{a}=50$~K.  The sound speed in the ambient gas is
$a_{a}=1.17$~km~s$^{-1}$.   Note that with these choices, the jet is
overpressured with respect to the ambient medium, i.e., $p_{a}\equiv
p_0/\alpha $ and $\alpha =200$.  Several purely adiabatic simulations
were performed for comparison with the cooling models, in this case we
have taken $p_{a}=p_0$, $T_a = 10,000$~K and $a_{a}=16.6$~km~s$^{-1}$,
and the jet is initially in pressure equilibrium with the ambient
medium.  Physically our radiatively cooling models
represent a dense jet beam surrounded by a hotter cocoon embedded in a
cold ISM.  The kinetic model of a dense jet embedded in a less dense
cocoon which is in turn ensheathed by a dense and cold molecular gas is
consistent with detailed observation of a number of protostellar jet
systems including HH111 (Nagar et al.\ 1997).

Since the strength and orientation of the magnetic field in
protostellar jets is not determined observationally, there is
considerable freedom in specifying the initial field configuration.  We
choose to focus on primarily toroidal fields in this paper (although we
do present models which include both toroidal and axial fields) as
theoretical models of magnetically driven outflows predict the field
should become primarily toroidal asymptotically.  In the absence of a
definitive theory for the radial profile of the magnetic field, we
assume the toroidal magnetic field is zero on the jet axis, achieves a
maximum at some radial position in the jet interior $r_m$, and returns
to zero at the outer jet boundary.  This implies that all currents
which support the field are confined within the jet (although as the
simulation progresses magnetic field, and therefore current, is
convected into the cocoon).  Within the jet, we adopt the simple
profile for the toroidal magnetic field
\begin{equation}
B_\phi (r)=\left\{ \begin{array}{cc} 
  B_{\phi ,m}\frac r{r_m} & 0\leq r\leq r_m \\
  B_{\phi ,m}\frac{R_j-r}{R_j-r_m} & r_m\leq r\leq R_j \\
  0 & R_{j}<r \end{array} \right.
\end{equation}
This profile is identical to that used by Lind et al.\ (1989) and by
FRJN-C when $0\leq r\leq r_m$ and when $r>R_j$ but
differs for $r_m\leq r\leq R_j$ where they used a force-free $B_\phi
(r)=B_{\phi ,m}(r_m/r)$ (which requires a large return current located at
the jet surface).  CG have studied both longitudinal and helical fields
of the form used by Todo et al. (1992) throughout the computational
domain.  Our toroidal magnetic field corresponds to a uniform
current density inside radius $r_m$, and a distributed return current
in the outer portion of the jet.  In adiabatic and cooling jet
simulations designed to be compared directly, the toroidal magnetic
field reaches a maximum value at $r_m/R_j=0.9$ with $\beta _1\equiv
\beta _m=8\pi p_0/B_{\phi ,m}^2=1$ ($B_{\phi ,m}=58.8$ $\mu G$) and
with $\beta _{1/4}\equiv \beta _m=0.25$ ($B_{\phi ,m}=117.6$ $\mu G$).
To illustrate the toroidal magnetic field profile adopted here, we plot
equation (5) for $\beta_m=1$ and $r_m=0.9$ in Figure 1.

The equation of hydromagnetic equilibrium
\begin{equation}
\frac d{dr}\left( p_{j}(r)+\frac{B_z^2(r)}{8\pi }+\frac{B_\phi ^2(r)}{8\pi }
\right) =-\frac{B_\phi ^2(r)}{4\pi r} , 
\end{equation}
has been used to establish a suitable radial gas pressure profile
in the jet
\begin{equation}
p_{j}(r)=\left\{ 
\begin{array}{cc}
\left\{ 2\left[ 1-(r/r_m)^2\right] \frac{B_{\phi ,m}^2}{8\pi p_0}%
+p_m/p_0\right\} p_0 & 0\leq r\leq r_m \\ 
\left\{ 1-\frac 2{(1-r_m/R_j)^2}\left[ 3(1-r/R_j)-(1-r^2/R_j^2)+\ln
(r/R_j)\right] \frac{B_{\phi ,m}^2}{8\pi p_0}\right\} p_0 & r_m\leq r\leq
R_j \\ 
p_0 & r=R_j 
\end{array}
\right. 
\end{equation}
where $p_m = p_{j}(r_{m})$ and
\begin{equation}
p_{j}(r_m)=p_0-\frac 2{(1-r_m/R_j)^2}\left[
3(1-r_m/R_j)-(1-r_m^2/R_j^2)+\ln (r_m/R_j)\right] \frac{B_{\phi ,m}^2}{8\pi}
 .
\end{equation}
The radial pressure profile which gives a
magnetostatic equilibrium is shown in Figure 1 where we plot equation
(7) for $\beta_m=1$ and $r_m=0.9$.  Since the jet density is held
constant, $T_{j}$ has the same functional form as $p_{j}$.  In some of
our simulations with strong toroidal fields, the condition of
magnetohydrostatic equilibrium within the jet beam requires negative
pressures in the neighborhood of $r_m$.  In these cases we simply fix
$p_m=0$, and scale the pressure profile given by equation (7) by an
appropriate amount so that this condition is met.  Of course, this
means that the jet beam is no longer in strict radial equilibrium.

The profile adopted for the magnetic field is an important
difference between this work and previous studies (FRJN-C, CG).  To
allow us to study the effect of the field geometry, we have also
performed a few simulations in which the field contains a uniform axial
component as well as the toroidal distribution given above, making the
field helical.  We will also describe the evolution of force-free
fields with $B_{\phi} \propto 1/r$.  We find that the detailed
structure of a magnetized radiative jet is sensitive to the precise
field geometry, thus uncertainty about this geometry is a serious
limitation to modeling observed systems.

To construct models of pulsed jets, we apply a purely
sinusoidal perturbation in time so the axial component of the velocity is
\begin{equation}
u_z (x=0) =  v_{j}(1 + A \sin \omega t)~,
\end{equation}
where we choose $A=0.25$, and for all of the simulations
$\omega=25a_0/R_j$ where $a_0=16.6$~km~s$^{-1}$ is a fiducial reference
speed (equivalent to the sound speed at a temperature of $10^{4}$~K).

The important dimensionless parameters which characterize the evolution
of the jet are the density ratio $\eta$ (taken to be 10 for all models
presented here), and the internal magnetosonic Mach number $M_{j,ms} =
v_{j}/a_{ms}$, where $a_{ms} \equiv (a_j^2+V_A^2)^{1/2}$ and $V_A$ is
the Alfv\'en speed.  In simulations without toroidal magnetic field or
with force-free toroidal magnetic field the pressure, density, and
temperature are uniform across the jet, so that the internal sonic Mach
number $M_{j} = v_j/a_j$ is constant.  However, a radially varying
magnetic pressure requires a radially varying thermal pressure in order
to initialize a jet beam in radial force balance and the sonic and
magnetosonic Mach numbers are not constant.  In general, we
characterize all our models by the magnetosonic Mach number on the jet
axis. However, since internal dynamics and timescales involve wave
propagation across the jet through a medium of varying magnetosonic
speed we also define linear radial averages as, for example,
\begin{equation}
\bar{M}_{j,ms}\equiv \frac 1{R_{j}}\int_0^{R_{j}}M_{j,ms}(r)dr . 
\end{equation}

\subsection{Tests of the Method}

Tests of the MHD algorithms used here have been reported in a number of
papers (Stone \& Norman 1992a, 1992b; Stone et al.\ 1992; Hawley \&
Stone 1995).  We have also confirmed that we are able to reproduce the
results of previous MHD studies of jets.  In particular, we have
performed simulations of adiabatic, underdense, magnetized jets using
parameters identical to Clarke, Norman, \& Burns (1986); we find excellent
agreement between our results with those reported by these authors.  In
particular, we find production of a nose cone with a similar aspect
ratio, strong pinch modes within the jet beam due to lack of initial
radial equilibrium, and strong suppression of vortical motion in the
cocoon by magnetic stresses.

\section{Jet stability}

Before describing the results of our simulations, let us review the
basic stability properties of supermagnetosonic jets in order to aid
interpretation.  The presence of toroidal magnetic fields and the
poloidal current can give rise to current driven pinch modes in
addition to velocity shear driven Kelvin-Helmholtz pinch modes. In the
absence of any zero order velocity shear, e.g., well inside the
velocity shear surface at the jet boundary, Begelman (1998) has shown
that a beam containing a toroidal magnetic field can be unstable to
current driven pinch modes when
\begin{equation}
\frac{d\ln B}{d\ln r}>\frac{\Gamma \beta (r)-2}{\Gamma \beta (r)+2} 
\end{equation}
where the plasma beta $\beta (r)\equiv 8\pi p(r)/B^2(r)$, and $\Gamma$ is
the adiabatic index.  In general, shorter pinch lengths will grow
faster.  If the beam contains a toroidal magnetic field and a weaker
uniform axial magnetic field the condition becomes
\begin{equation}
\frac{d\ln B}{d\ln r}>-1+\frac{1}{2}(kr\frac{B_z}{B_\phi(r)})^2 ,
\end{equation}
where $k$ is the wavenumber of the mode;  in general, shorter pinch
lengths, i.e., larger $k$, are stabilized. 

The velocity driven Kelvin-Helmholtz unstable pinch body modes might
appear near resonances which, in the absence of radial variation in
parameters in the jet and in the external medium, are given by (e.g.,
Hardee, Clarke, \& Rosen 1997; Hardee \& Stone 1997)
\begin{equation}
\eqnum{12a} 
\omega R_{j}/a_{c}\approx (2m+1/2)\pi /2
\end{equation}
\begin{equation}
\eqnum{12b} 
\lambda /R_{j}\approx \frac{2\pi }{\omega R_{j}/a_{c}}\frac{M_{j,ms}}{
1+M_{j,ms}/M_{c}} 
\end{equation}
where $\omega$, $m$, and $\lambda$ are the angular frequency, body mode
number $m =$ 1, 2, 3, etc., and wavelength, respectively.  In the
expressions above $M_{c}\equiv v_{j}/a_{c}$ and $a_{c}$ is the
sound speed in the cocoon medium immediately outside the jet and not
the sound speed in the undisturbed ambient medium.  At these resonances the
pinch modes grow with spatial growth
rate
\begin{equation}
\eqnum{13}
k_I\approx -(2M_{j,ms}R_{j})^{-1}\ln \left( 4\omega R_{j}/a_{c}\right)  ,
\end{equation}
where $k_I$ is the imaginary part of the wavenumber.  The spatial
growth length over which a linear perturbation increases in amplitude
by a factor $e$ is $\ell =\left| k_I^{-1}\right| \approx
M_{j,ms}R_{j}$.  On the high Mach number axisymmetric flows that we
simulate here these Kelvin-Helmholtz modes grow relatively slowly.

\section{Steady Jet Simulations}

Simulation parameters for the steady jet models discussed here
are listed in Table 1.   In what follows, we
report times in units of a sound crossing time of the jet beam, $\tau_{0}
\equiv R_j/a_j \approx 152$~yr where $a_j = 5.25$~km~s$^{-1}$.
Magnetosonic crossing times of the jet beam are in the range, $\tau
_{ms} \equiv R_j/\bar{a}_{ms} \approx 150 - 58$~yr where $\bar{a}_{ms}$
is the linear average of the magnetosonic speed.  A grid crossing time
can be defined by $\tau _g\equiv L_{grid}/v_h\approx 314$~yr, where
$v_h=\left\{ \eta ^{1/2}/\left[ 1+ \eta ^{1/2}\right] \right\} v_{j}
\approx 250$~km~s$^{-1}$ is the velocity of advance of the head of a
cold (thermal and magnetic pressure much less than the ram pressure)
jet.  We stop the evolution when the bow shock has reached $\sim
100R_j$.  The spatial evolution of the jet's transverse structure is then
evaluated at an axial distance of 68.5R$_j$ (a location behind the
region strongly affected by jet head vortices), and the spatial evolution
of the cocoon's structure can be evaluated through comparison of cocoon
properties at this distance and near to the inlet.

\subsection{Magnetized Adiabatic Jets}

In order to understand the dynamics of our magnetized radiative jets,
it is necessary to first compute non-radiative jet models with an
identical magnetic field profile.
We have computed three models:  a purely
hydrodynamical jet ($\beta_{\infty}^{Ad}$), a weak field jet ($\beta
_{1}^{Ad}$), and a strong field jet ($\beta _{1/4}^{Ad}$).  Table 1
gives additional parameter values adopted for each simulation.  Unlike
our radiatively cooling jet models, all three adiabatic models are
initially in pressure equilibrium with the ambient medium; i.e. we
adopt $T_a=10^{4}$~K rather than the $T_a=50$~K used for the cooling
jets.  Figure 2 plots the logarithm of the density at the end of each
of these three simulations.  Figure 3 plots the toroidal magnetic field
strength $B_{\phi}$ at the same times for the two magnetized jet
models.  Note only the final $50R_j$ of the computational domain is
shown in both figures for clarity.

Comparison of the final structure of each jet shows remarkably little
difference between these models.  In the purely hydrodynamical jet
simulation $\beta _{\infty}^{Ad}$ the jet remains well collimated
within its initial radius inside a low density cocoon of typical
density one-half the ambient and one-twentieth the jet density, i.e.,
n$_{c} \approx 50$ cm$^{-3}$, and has a typical radius $ \approx
10R_j$.

The weak toroidally magnetized jet simulation, $\beta _{1}^{Ad}$,
experiences some slight collimation from the toroidal magnetic field
so that by
an axial distance of 68.5 R$_j$, where r$_{j} \lesssim R_j$, the
toroidal field maximum has increased by about 7\% as it has moved
inwards with the contracted jet.  However, unlike the $\beta _\infty^{Ad}$
jet we find three significant pressure pulses with amplitudes up to
twice p$_j(r)$ and spacing of about 15R$_j$ (with the first pulse at an
axial distance of $\approx 30R_j$) in the jet interior when
$0.1<r/r_{j}<0.6$.  The amplitudes are largest at $r\approx 0.3r_{j}$.  This
location is not consistent with pressure perturbations produced by the
Kelvin-Helmholtz instability at the jet surface. The confinement of
these pressure perturbations to the jet interior is consistent with a
current driven instability, moreover this region satisfies the
condition for instability (eq.\ [11]) found by Begelman (1998).

The strong toroidally magnetized jet simulation, $\beta _{1/4}^{Ad}$,
cannot be introduced at the inlet with a true equilibrium pressure
distribution: this would require a negative minimum jet pressure. As a
result the jet spine expands to $r_{j}\approx 1.5R_j$ at an axial
distance of $68.5R_j$ and then recollimates somewhat under the
influence of higher external pressure near to the jet head and the
stronger toroidal magnetic field.  At axial distances between
$80-95R_j$ there is evidence for pressure pulses associated with
pinching in the jet interior at $r<0.2r_{j}$. Three main pressure
pulses with spacing of $\sim 6R_j$ and substructure at one-third this
spacing might be due to current driven instability.  Neither of the
toroidally magnetized adiabatic jet simulations show evidence for a
magnetically confined ``nose cone'' (Clarke, Norman, \& Burns 1986;
Lind et al.\ 1989).

\subsection{Magnetized Cooling Jets}

Images from three cooling jet simulations, a purely hydrodynamical
model ($\beta _{\infty}^{Cl}$), a weak field model ($\beta _{1}^{Cl}$),
and a strong field model ($\beta _{1/4}^{Cl}$) with magnetic field
configuration identical to that used in the three adiabatic jet
simulations described in \S4.1 are shown in Figure 4. Figure 5 shows
the toroidal magnetic field strength $B_{\phi}$ at the same times for
the two magnetized jet models.  Note only the final $50R_j$ of the
computational domain is shown in both cases. The purely hydrodynamical
jet simulation $\beta _{\infty}^{Cl}$ is directly comparable to the
results presented in Stone \& Norman (1993; 1994); the primary difference
being the use here of the coronal cooling curve of Dalgarno \& McCray
(1972), as opposed to the non-equilibrium cooling rates used
previously.

Unlike the adiabatic jets discussed above, the addition of a toroidal
magnetic field to radiatively cooling jets leads to noticeable
differences relative to the non-magnetized $\beta_\infty ^{Cl}$
simulation.  For example, in $\beta_1^{Cl}$ the jet beam has expanded
in width, and the cocoon is much more uniform and smaller relative to
$\beta_\infty ^{Cl}$.  Jet expansion in $\beta_1^{Cl}$ has resulted in
a 50\% decrease in the toroidal magnetic field strength in the jet
beam.  No nose cone is evident in either the $\beta_\infty ^{Cl}$ or
$\beta_1^{Cl}$ simulations.

The $\beta _{1/4}^{Cl}$ simulation is strongly influenced by expansion
of the jet resulting from the non-equilibrium initial configuration at
the inlet.  Figure 6 plots radial profiles of the density, axial
velocity, temperature, and toroidal magnetic field strength at an axial
distance of 68.5 R$_j$ from the jet nozzle in all three simulations.
From the figure, the structure of the $\beta _{1/4}^{Cl}$ jet can be
seen to consist of an expanded spine with r$_{j}\approx $ 3.5R$_j$
surrounded by a sheath of thickness $\approx $ 3.5R$_j$. The magnetic
profile shows an approximate linear increase from the axis to 2R$_j$
and plateau to 4R$_j$ with B$_\phi \approx 21$ $\mu G$.  At the same
time, the jet temperature drops from 2,000 K on axis to a low of $\sim
$ 100 K in the outer portion of the jet, and then jumps up to $\sim $
2,000 K in the cocoon.

The stronger magnetic field along with jet expansion have resulted in
significant morphological changes in jet and cocoon structure beyond an
axial distance of about 70 R$_j$ in $\beta_{1/4}^{Cl}$.  At this
distance the jet velocity suddenly decreases by more than a factor of
two at an ``upstream shock'' and a nose cone of length $\approx 25R_j$
is ``pushed'' ahead of a ``downstream shock''. The nose cone drives a
more conventional jet terminal and bow shock.  The temporal evolution
leading to this structure is shown in Figure 7, which plots the
logarithm of the density at 5 times during the simulation, $(0.4, 0.8,
1.2, 1.6, 1.8) \times \tau_0 \approx (61, 122, 182, 243, 274)$~yr.
Initially the jet remains well collimated.  Strong cooling at the head
of the jet results in most of the shocked ambient and jet gas
collecting in a dense shell ahead of the Mach disk.  Once the jet has
propagated about $20R_j$, it begins to expand due to lack of
magnetohydrostatic balance.  Consequently, the head of the jet grows
laterally, and becomes filamentary.  Most of the shock processed gas is
confined to a loosely defined ``nose cone" ahead of the Mach disk.
Detailed structure at the head of the jet is shown further in Figure 8,
which plots the logarithm of the temperature, the divergence of the
velocity (an indicator of shocks), and velocity vectors at the end of
the simulation.  Comparison of the temperature and velocity divergence
plots clearly shows that the highest temperatures are located in
filaments immediately behind strong shocks in the flow, such as the
Mach disk and outer bow shock.  Very little non-axial motion is
observed in the velocity vectors.

\subsection{Cooling Jets with Different Magnetic Field Profiles}

Three additional magnetic field profiles have been studied for steady
radiatively cooling jets. One simulation, $\beta _{ff}^{Cl}$, has a
toroidal force-free field in the jet with $B_\phi \propto 1/r$. In this
simulation the maximum field strength $B_{\phi ,m}=236.3$ $\mu G$
($\beta _m=0.062$) at $r_m/R_j=0.05$ (the innermost computational zone)
and the magnetic field is set to zero at the jet surface. Within the
force-free jet, temperature and thermal pressure are constant. In a
second simulation, $ \beta _{Rm}^{Cl}$, the magnetic and pressure
profiles given by equations (5) and (7) are used with $\beta _m=0.25$
($B_{\phi ,m}=117.7$ $\mu G$) at $r_m/R_j=0.2$. Unlike the $\beta
_{1/4}^{Cl}$ simulation, in this simulation a true equilibrium pressure
profile can be achieved and outside the toroidal field maximum the
field configuration is not too different from the force free simulation
$\beta _{ff}^{Cl}$.  Finally, a third simulation, $\beta _{\phi
z}^{Cl}$, includes an axial magnetic field along with a toroidal
magnetic field. In this simulation a constant axial magnetic field in
the jet and in the ambient medium of strength $B_z=58.8$ $\mu G$
($\beta _z=1$) has been added to a toroidal magnetic field profile
given by equation (5)
with, $ r_m/R_j=0.8$ and $\beta_m=0.25$ ($B_{\phi ,m}=117.7$ $\mu G$).
In this simulation, the pressure profile in the jet is similar to that
in the $ \beta _{1/4}^{Cl}$ simulation and is not in equilibrium.
Figure 9 plots the logarithm of the density at the final time for each
of these simulations (along with the $\beta _{1/4}^{Cl}$ simulation shown
in Figures 7 \& 8), and Figure 10 plots the corresponding toroidal
magnetic field strength.  Radial profiles of a variety of quantities
in the jet beam for each simulation
at an axial distance of 68.5R$_{j}$ are given in Figure 6.

In $\beta _{ff}^{Cl}$ the jet cools radiatively and the axial velocity
remains constant out to $\sim $ 90 R$_j$ where the velocity abruptly
drops by about 20\% at a terminal ``upstream'' shock. A modest
nose cone of material about 9 R$_j$ in length precedes a ``downstream''
shock and drives a terminal and bow shock. The jet expands as the
initial jet pressure is above the radiatively cooled cocoon pressure at
the inlet.  The radial structure (Figure 6) is
similar in dimension
to that found in the $\beta _\infty ^{Cl}$ and $\beta _1^{Cl}$
simulations,
although the cocoon size is slightly less than in
the $\beta _1^{Cl}$ simulation.

The $\beta _{Rm}^{Cl}$ simulation serves as an intermediate case
between the force-free $\beta _{ff}^{Cl}$ simulation and the non
force-free $\beta _{1/4}^{Cl}$ simulation, and can also be compared to
the $\beta _1^{Cl}$ simulation. Inside the jet when  $r>0.2r_j$ the
magnetic field profile is very similar to the force-free case. At the
inlet the jet is overpressured relative to the radiatively cooled
cocoon but jet expansion does not begin until an axial distance of
about 40~R$_j$. Interior to this distance nine well defined pressure
oscillations with a spacing of $\sim $ 3.3~R$_j$ are observed at jet
radii $r<0.15r_j$, and probably are the result of a current driven
pinch mode.  Pressure fluctuations attributable to this pinch mode are
on the order of $\pm~2\%-3\%$ and these fluctuations disappear as the
jet begins to expand. At an axial distance of 68.5 R$_j$ (Figure 6)
the jet spine and sheath are reduced in size by about 25\% relative to
the force-free configuration.  The jet is in approximate pressure
balance with the cocoon at this point.  We note that the jet spine
contains a relatively high pressure and temperature central region
confined by the magnetic field.  The cocoon is also reduced in size by
about 25\% relative to the force-free configuration.  A modest nose
cone about 10 R$_j$ long leads the jet in this case after a velocity
drop of 15\% in an ``upstream'' shock.

Finally in the $\beta_{\phi z}^{Cl}$ simulation we consider the effect
of adding an axial magnetic field to the toroidal magnetic field
configuration used in $\beta_{1/4}^{Cl}$. The axial field imposed over
the entire computational grid has served to ameliorate the large jet
expansion observed in $\beta _{1/4}^{Cl}$ and at an axial distance of
68.5 R$_j$ (see Figure 6) the jet spine r$_{j}\approx 3R_j$ and sheath
of thickness $ \approx R_j$ together equal the jet spine width in
$\beta _{1/4}^{Cl}$.  This width is similar to that seen for the $\beta
_{ff}^{Cl}$ jet and in this simulation the jet is at lower pressure
than the hotter cocoon.  Toroidal magnetic field resides within the jet
spine and sheath, while the axial magnetic field has been reduced within the jet
by jet expansion but increased by compression in the outer part of the
cocoon.  This additional magnetic field and accompanying high sheath
and higher cocoon density has helped to confine the jet.  The presence
of axial magnetic field has almost but not quite eliminated the
extensive nose cone observed in the $ \beta _{1/4}^{Cl}$ simulation.

\section{Pulsed Jets}

Some of the most prominent structural features in highly collimated
jets associated with protostellar objects are the bright emission knots
in the flow close to the source (see Reipurth 1997 for a recent review
of the observations).  It has been suggested that symmetric pinch modes
of the Kelvin-Helmholtz (K-H) instability might give rise to such knots
(B\"{u}hrke, Mundt, \& Ray 1988), which has motivated detailed hydrodynamical
modeling of the nonlinear evolution of symmetric K-H modes in cooling
jets (Massaglia et al.\ 1992; Bodo et al.\ 1994; Rossi et al.\ 1997;
Downes \& Ray 1998).  However the large proper motion observed in some
cases (Eisl\"{o}ffel \& Mundt 1994), and direct kinematic evidence
provided by spectroscopic observations (Reipurth 1997) have shown that
some knots are associated with velocity variability in the outflow.
Thus, while the nonlinear stage of {\em nonaxisymmetric} modes of the
K-H instability might still prove relevant to understanding the helical
structures or wiggles observed in some protostellar jets (Hardee \&
Stone 1997; Stone, Xu, \& Hardee 1997), most of the internal knot
structure in the jets is probably produced by flow variability in the
jet beam.

The hydrodynamics of velocity-variable, or pulsed, jets has been widely
studied in the literature (see Raga 1993 for a review).  In an
important early contribution, Raga \& Kofman (1992) demonstrated that a
sinusoidal velocity variation at the jet inlet steepens into a sawtooth
pattern downstream.  One-dimensional (Hartigan \& Raymond 1993),
followed by multidimensional hydrodynamical simulations (Stone \&
Norman 1993; de Gouveia Dal Pino \& Benz 1994; Biro \& Raga 1994; Biro
1996; Suttner et al.\ 1997) have shown that the velocity
discontinuities in the sawtooth pattern consist of a shock pair: an
upstream shock decelerating high velocity gas as it collides with the
pulse, and a downstream shock sweeping up low velocity material ahead
of the pulse.  These shock pairs move apart at a constant rate causing
the pulse width to grow linearly in time up to some asymptotic value.
In the absence of magnetic fields this evolution can be understood by
a simple analytic model (Falle \& Raga 1993).  Recently, there has been
considerable effort in understanding how such ``internal working
surfaces'' (Raga et al.\ 1990) affect the dynamics and observed
properties of jets.
In this section, we present the results of an investigation of the magnetohydrodynamics (MHD) of
radiatively cooling pulsed jets.

Parameters for the simulations discussed below are listed in Table 2.
Because shock compression on the downstream and
upstream sides of pulses generated by velocity fluctuation is a
function of jet radius, we include in Table 2 area weighted averages
as, for example,
\begin{equation}
\langle{M}_{j,ms}\rangle \equiv \frac {2}{ R_{j}^2}\int_0^{R_{j}}M_{j,ms}(r)rdr\text{ .} 
\end{equation}

\subsection{Variation in Magnetic Field Strength}

We have computed a
purely hydrodynamic model with $\beta_{\phi} = \infty$, hereafter
denoted as $\beta_{\infty}^{P}$, to serve as a benchmark.  Two simulations
were performed using a toroidal magnetic field profile in the jet beam
given by equation 5 in Paper II, with r$_m = 0.9R_j$.  The first model
has a peak field strength of $B_{\phi,m}=58.8 \mu G$ corresponding to
an equipartition field, i.e.  $\beta_{\phi,m} \equiv 8\pi
p_0/B_{\phi,m}^2 = 1$, hereafter denoted as $\beta_{1}^{P}$.  The second
model has a peak field strength $B_{\phi,m}=117.6 \mu G$ corresponding
to a strong field where $\beta_{\phi,m} = 0.25$, hereafter
denoted as $\beta_{1/4}^{P}$.  These profiles are identical to those adopted
in the steady jet simulations $\beta_{\infty}^{Cl}$, $\beta_{1}^{Cl}$, and
$\beta_{1/4}^{Cl}$.

In Figure 11 we show images of the logarithm of the density at the final
time in the three simulations.  The $\beta_{\infty}^{P}$ model is directly
comparable to previous results of hydrodynamical simulations, e.g.,
Stone \& Norman (1993).  Despite the use of a different numerical method
(a PPM algorithm was used in this earlier work, instead of the ZEUS
code used here) and
lower numerical resolution, there is excellent agreement in the overall
structure of the jet.  The steepening of the sinusoidal pulses into
thin, dense sheets by $z \approx 10R_j$ is evident.  Thereafter, the
pulses are bounded by two shocks.  Significant mass flux into the
cocoon from the high pressure shocked gas within the pulse is evident
as smooth flows.  Each pulse is seen to drive a shock into the cocoon,
so that the cocoon is much broader than steady jet models using the
same parameters.  The structure at the head of the jet is complex as
pulses begin to merge with one another, and interact with the ambient
medium.  Note that the pulse width increases linearly with distance
from the nozzle.  For the purely
hydrodynamic jet, the rate of increase in size is set by the balance
between radial mass loss into the cocoon from the pulse ``surface''
and the mass flux into the pulse
through isothermal shocks (Falle \& Raga 1993).

As the magnetic field strength is increased (models $\beta_{1}^{P}$ and
$\beta_{1/4}^{P}$ shown in the lower two panels of Figure 11), the gross
properties of the pulsed jet remain the same.  Two features of the
pulses are clearly modified by the magnetic field.  Firstly, the
postshock density within the pulses on the axis increases with the
field strength.  Secondly, the rate of increase in the pulse width
increases with increasing magnetic field strength. While the presence
of hoop stresses associated with the toroidal field clearly influences
the amount of material ejected from the pulses into the cocoon, the
maximum radial distance that material is ejected into the cocoon does
not decrease with field strength.  In all three simulations, the wings
generated by the pulses extend to a radial distance of $\sim 5R_j$.
Figure 12 plots the logarithm of the toroidal magnetic field strength in
$\beta_{1}^{P}$ and $\beta_{1/4}^{P}$.  The field acts as a tracer of jet
material since the ambient gas is unmagnetized.  Compression of the
magnetic field by the internal shocks associated with the pulses, and
ejection of material and the magnetic field into the cocoon is
evident.  Note this ejection
occurs over a larger area in $\beta_{1/4}^{P}$ as might be
expected given the larger pulse width. 

In order to illustrate the variation of properties of the pulses with
magnetic field strength, we plot in Figure 13 the density, pressure,
axial velocity, and toroidal magnetic field strength along an axial
slice through the center of all three jets at time $t=1.2 \tau_0$ --
note that this time is not the same as that used in Figures 11 and 12.
Only a limited segment of the axial domain between $20R_{j}$ and
$45R_{j}$ is shown for clarity.  The plot shows three pulses developing
within this region, one centered near $23R_j$, the next at $31R_j$, and
the last at $38R_j$.  In both cases the magnetic field strength
declines outwards from the origin over the region shown in Figure 13 by
a factor 2 -- 3.  Several trends are clearly evident in the plot.  Note
the increase in width of the pulses with axial distance.  The strongly
magnetized model shows the widest pulses.  Interestingly, the gas
pressure is nearly identical within all three jets.  The axial density
is much higher in the strongly magnetized pulsed jet, increasing from
5 times (in the first pulse) to 20 times (in the last pulse)
larger than the hydrodynamic pulsed jet.
The axial velocity shows the classic sawtooth
pattern expected for a sinusoidally pulsed jet (Raga \& Kofman 1992),
modified by the shock pair which appears at each step (e.g., Stone \&
Norman 1993).  The toroidal field strength follows the density profile
closely as would be expected in ideal MHD when the magnetic field is
parallel to the shock interface.

In all three jets, the pulse width increases linearly with axial
distance.  At $z \sim 50 R_j$, the pulse widths are about $4R_j$,
$5R_j$, and $7R_j$ in $\beta_{\infty}^{P}$, $\beta_{1}^{P}$,
and $\beta_{1/4}^{P}$, respectively.  To zeroth order the linear growth rate
in the pulse width scales with $1/\langle{M}_{j,ms}\rangle$ rather than
the sound speed (Falle \& Raga 1993) (see Table
2).  In $\beta_{1}^{P}$ and $\beta_{\infty}^{P}$ the pulses remain separated
along the entire length of the jet, so that the upstream and downstream
shock pair is evident nearly to the jet head.  On the other hand, the
spreading of the pulse width is so large in the $\beta_{1/4}^{P}$ model that
the pulses merge before reaching the head of the jet, leading to the
disappearance of the upstream shock -- leaving only a sequence of
downstream shocks --  at $z > 60R_j$.

The observed differences in pulse density, radial density profile and
pulse width are the result of hoop stresses which confine some of the
shocked jet material within the pulses to near the axis, and the result
of the lack of radial pressure equilibrium in the material
after it passes through the upstream and downstream shocks into the
pulse.  Recall that the unshocked material in $\beta_{1}^{P}$ is in radial
pressure equilibrium.  In an isothermal shock
the toroidal magnetic pressure increases $\propto \rho^2$ whereas
thermal pressure increases $\propto \rho$.  Thus, in general, the
toroidal plasma $\beta$ behind the shocks should decrease relative to the
toroidal plasma $\beta$ in the unshocked jet material by up to $\propto
\rho^{-1}$.  Of course, the toroidal plasma $\beta$ in the unshocked and
shocked pulse material is a function of radius and the relative
importance of thermal versus magnetic pressure
is further modified by material and magnetic ejection from
the pulse.  However, in $\beta_{1/4}^{P}$ the area weighted plasma $\beta$ in a
typical pulse is $\langle\beta_{\phi}\rangle \lesssim 2$ (compared to
$\langle\beta_{\phi}\rangle \sim 8$ in the unshocked jet material), and
in this case magnetic pressure can play a significant role in supporting
the pulses against ram pressure.  

Considerable toroidal field accumulates at the head of the jet into a
nose-cone-like structure; this is most evident in the very strong field
case.  For our parameters typical magnetic field strengths in the pulses
at $z > 20R_j$ are about $50 \mu$G and $75 \mu$G in $\beta_{1}^{P}$ and
$\beta_{1/4}^{P}$, respectively, and the field strength declines as the
distance from the inlet increases.  Between the pulses the magnetic
field strength is reduced from these values by over a factor of 10.
Near the head of the jet in the complex
interactions between shocked jet material ejected from the pulses and
shocked ambient gas, field strengths are as high as $190 \mu G$ in
$\beta_{1}^{P}$, and $570 \mu G$ in $\beta_{1/4}^{P}$.

\subsection{Radial Variation in Magnetic Field and Pulse Structure}

We have studied the effect of varying the geometry of the magnetic
field in the jet beam using three different simulations.  The first,
$\beta_{1}^{P}$, is the model discussed above in which $B_{\phi,m}=58.8 \mu
G$ at $r=0.9R_{j}$.  The second, $\beta_{Rm}^{P}$, is another equipartition
toroidal magnetic field model but with $B_{\phi,m}=58.8 \mu G$ at
$r=0.2R_{j}$.  The third, $\beta_{\phi z}^{P}$, is a model with
$B_{\phi,m}=58.8 \mu G$ at $r=0.8R_{j}$, and which contains a uniform
axial field of $B_{z}=58.8 \mu G$, with $\beta_z \equiv 8\pi p_0/B_z^2
= 1$ added throughout the computational domain.  In all three of these
simulations, the jet beam is initially in exact radial equilibrium.

Images of the logarithm of the density at the final time in each of
these simulations are shown in Figure 14.  A comparison between
$\beta_{1}^{P}$ and $\beta_{Rm}^{P}$ shows that when the magnetic field is
peaked close to the axis, the pulses widen non-uniformly, growing in
width more rapidly near the axis of the jet than at the surface.  The
size and shape of the wings of the pulses is similar in both models.  A
strong axial magnetic field (model $\beta_{\phi z}^{P}$) helps to confine
material in the pulses and reduces the amount of material ejected to
the cocoon.

Images of the logarithm of the toroidal magnetic field strength in
$\beta_{1}^{P}$, $\beta_{Rm}^{P}$, and $\beta_{\phi z}^{P}$ are shown
in Figure 15.
In $\beta_{1}^{P}$, considerable toroidal magnetic field has been injected
into the cocoon.  Since the field tracks jet material, this indicates a
large amount of shocked jet gas is ejected from the pulses.  On the
other hand, in $\beta_{Rm}^{P}$ the strongest toroidal field is found close
to the axis, and there is much less magnetic field in the cocoon,
indicating reduced ejection of shocked jet material from the pulses.
In $\beta_{\phi z}^{P}$ the toroidal magnetic field is more confined
towards the axis indicating that considerably less material is ejected
from the pulses into the cocoon, and the cocoon is smaller in this
simulation than in the other two simulations.

In Figure 16 we plot various quantities as a function of radius for
pulses at an axial distance $z/R_j =$ 48, 53, 51, 47.5 and 51 at the
final time in simulations $\beta_{\infty}^{P}$, $\beta_{1}^{P}$,
$\beta_{1/4}^{P}$, $\beta_{Rm}^{P}$ and $\beta_{\phi z}^{P}$,
respectively.  These locations are the centers of last pulse which is
non-interacting with its neighbors at the final time in each
simulation.  Note that $\beta_{1/4}^{P}$ and $\beta_{Rm}^{P}$ result in
similar axially peaked density pulse profiles (although the axial
density in $\beta_{1/4}^{P}$ is four times that in $\beta_{Rm}^{P}$).
Other magnetic field configurations are less axially peaked but are
still very different from the ``top hat'' profile of the hydrodynamical
simulations (solid line in each panel).  In all cases the magnetized
pulses are significantly colder on the axis, typically by more than a
factor of 2, than the hydrodynamic pulse.  The hydrodynamic pulse has a
``top hat'' temperature profile with $T_{pulse} \approx 1,300$~K.
Expansion velocities evaluated at $r=0.8R_j$ lie within $\pm 15\%$ of
$a_j$ ($a_j=a_0/3.16$) and are relatively independent of magnetic field
strength or configuration.  Thus, we might expect an equilibrium pulse
width (mass into the pulse through upstream and downstream shocks
balanced by transverse mass ejection) to scale approximately
proportional to the density at the $r=0.8R_j$ surface, i.e., with the
mass ejection rate per unit surface area evaluated at $r=0.8R_j$.

At least approximately we would expect the pulse expansion rate and
ultimate width in $\beta_{1}^{P}$ and $\beta_{\phi z}^{P}$ to be comparable but
twice that (and in
$\beta_{1/4}^{P}$ and $\beta_{Rm}^{P}$ comparable but about four times that)
observed in the hydrodynamic case.
Our simulations do in fact show an increase in the rate at
which pulses widen consistent with this analysis, although the rate
of this increase is not as fast as this simple argument would predict.

From Figure 16, the maximum radial ejection velocity from the pulses
occurs at a radial distance much larger than
the pulse surface at $r=0.8R_{j}$.  It is affected by
the magnetic field configuration and strength, with the highest
velocities observed for the strongest toroidal magnetic field case,
$\beta_{1/4}^{P}$, and lowest velocities observed when an axial magnetic
field is present, $\beta_{\phi z}^{P}$.
Changes to the unshocked toroidal plasma
$\beta$ in the pulses is also dependent on the magnetic field
configuration and strength, and the radial variation in $\beta_{\phi}$
shown in Figure 16 is much different from that at the inlet (see
Table 2).  If we characterize the effects of the magnetic field by the
area weighted toroidal plasma $\beta$ then we find that
$\langle\beta_{\phi}\rangle_{pulse} \approx
\langle\beta_{\phi}\rangle_{jet}$ in cases $\beta_{1}^{P}$ and $\beta_{\phi
z}^{P}$, whereas $\langle\beta_{\phi}\rangle_{pulse} <<
\langle\beta_{\phi}\rangle_{jet}$ in cases $\beta_{1/4}^{P}$ and
$\beta_{Rm}^{P}$.  Note that this properly groups the magnetic simulations
with their observed spreading rates. However, only in case $\beta_{1/4}^{P}$
where $\langle\beta_{\phi}\rangle_{pulse} \sim 1.5$ can there be
significant magnetic pressure effects.  We conclude that the
differences in pulse width spreading rate are the result of the radial
density profile in the pulse induced by radial pressure equilibrium
with $B_{\phi}$, and clearly the density profile is significantly
modified even when the area weighted toroidal plasma $\beta$ is much
greater than one.

\section{Summary}
\subsection{Steady Jets}

Although inclusion of a weak magnetic field (weak in the sense that
the flow is highly supermagnetosonic) has little effect on the propagation
of adiabatic, overdense jets (e.g. Figures 2 and 3), such fields have
strong effects on cooling, overdense jets (e.g. Figures 4, 5, 9 and 10).
The difference is caused by the much larger compression of jet
and ambient gas that occurs in cooling shocks.  This compression amplifies
the magnetic field until in some cases the magnetic pressure is comparable
to the gas pressure in shocked gas,
a situation which never occurs in the adiabatic jet simulations.

Our simulations reveal at least three effects of the magnetic field
on steady, cooling jets.  Firstly, the field affects the structure and
fragmentation of dense sheets of cooled gas (e.g. Figures 4 and 9).
This is in part because magnetic pressure can support cooling gas against
compression, and so limit the density and width of cooling shells.
Alternatively, magnetic stresses can inhibit the radial flow of gas
away from the axis, leading to higher densities in some cases.
Both of these effects are evident in comparison of models
$\beta_{\infty}^{Cl}$, $\beta_{1}^{Cl}$, and $\beta_{1/4}^{Cl}$ (Figure 4).
The structure of the magnetic field is important to its overall effect,
for example, helical magnetic fields appears to produce smoother
fluctuations in the cocoon than purely toroidal fields of the same strength
(see Figure 9).  Much of this may be attributed to the difference 
the field has on the radial structure of the jet, either through
expansion or compression of the jet beam by pressure and hoop stresses.
Jets which undergo radial expansion have a much larger cross section near their
tip, and drive larger, more fragmented shells into the ambient gas.
Jets which are confined or compressed by the magnetic field tend to
have narrower and smoother cocoons.

Secondly, the presence of a magnetic field can, in some cases, result
in the production of a dense, plug of shocked jet material ahead of the
Mach disk (a ``nose cone").  The formation of nose cones has been noted
previously in simulations of magnetized extragalactic jets (Clarke et
al. 1986).  However, we find the formation of nose cones in cooling
jets is sensitive to the geometry of the field: only purely toroidal fields
which peak near the surface of the jet form nose cones, simply because
this configuration maximizes the hoop stresses which confine the nose cone.
Moreover, nose cones are likely to unstable in 3D which will
limit their relevance to real protostellar jets.

Finally, the presence of magnetic fields affects the stability of the
jet beam.  We do not observe any evidence for the MHD K-H instability
in our simulations, nor do we expect to given that
the linear growth rates for such modes are small for the supermagnetosonic
jets studied here.  Interestingly, we observe axial pressure fluctuations
in a few simulations where the jet beam is unstable to current driven
pinch modes (Begelman 1998).  However, our simulations reveal such modes
saturate at low amplitude in the nonlinear regime, so that it is unlikely
they will have much relevance to the internal knots observed in most
protostellar jets.

\subsection{Pulsed Jets}

Our simulations reveal that weak magnetic fields have a number of
effects on time-variable (pulsed) cooling jets.  Such pulsing results
in the formation of dense knots of shocked jet material in the jet
beam.  The inclusion of a toroidal magnetic field affects the rate at
which such knots grow by inhibiting the radial flow of material out of
the knots and into the cocoon.  This results in much higher densities
in the pulses as the field strength is increased (Figure 13).  The {\em
increase} of the density in the cooling gas in the shocked pulses as
the magnetic field strength is increased contradicts the expectations
of simple planar shock theory: this result emphasizes the importance of
multidimensional effects in MHD flows.

A second effect of a non-uniform magnetic field is the
introduction of radial structure into the density and pressure 
of the pulses, even if the jet is uniform at the nozzle.  Jets with
strong toroidal fields have radial density profiles in the pulses which
are strongly peaked toward the axis.  This profile is introduced by
radial variations in the magnetic pressure and hoop stresses in the
shocked pulses, which confines material near the axis.  It is
unlikely that real protostellar jets have a uniform (``top-hat") profile:
it has been shown that the structure of the cocoon can be affected by
the profile of the density and velocity in the jet beam through
in hydrodynamical simulations (Suttner et al 1997).  Our results
reinforce the importance of understanding the radial profile of
the outflow to interpret the structures observed.

Finally, even though toroidal fields affect the internal radial structure
of the pulses, we do not observe much effect on the structure of the
cocoon due to the magnetic field.  Radial profiles of various
quantities in the pulses (Figure 16) show that the {\em maximum} axial
velocity of material ejected from the pulses into the cocoon is 
relatively insensitive to the field strength.  Thus, the size of the cocoon
is not significantly decreased in the magnetized jets.  The radial
ejection velocity is small (a few times the sound speed in the unperturbed
jet beam), and it would require a delicate force balance to
completely eliminate radial outflow.

\section{Conclusions}

We have studied the propagation of both steady and time-variable (``pulsed")
protostellar jets using numerical MHD simulations.  Although we focus
our attention on models in which the jet beam contains a purely
toroidal magnetic field peaked near its surface, we have also studied
the effect of varying both the strength and geometry of the field on
the dynamics.

We find that even a weak magnetic field ($B \leq 60 \mu$G) in the jet
beam can lead to important effects on the structure and dynamics of
steady jets (FRJN-C; CGH; CG).  In particular, such fields can alter the density
structure and fragmentation of dense shells formed in cooling jets.
However, the details of the effects depend sensitively on the geometry
of the field.  For example, while magnetic pressure can limit
compression in such shells, resulting in {\em lower} densities and less
fragmented shells, radial hoop stresses associated with purely toroidal
fields can confine shocked gas towards the axis, resulting in {\em
higher} densities there.  In some cases, hoop stresses lead to the
formation of nose cones ahead of the Mach disk.  We see no evidence for
significant structure induced by the Kelvin-Helmholtz instability (such
modes were seen for the parameters and field geometry
adopted by CG).  We do see evidence
for current driven pinch modes but the induced pressure pulses in the
jet interior do not appear strong enough to explain the emission knots
that are observed in protostellar jets (see also CG).  Thus, models invoking jet
velocity fluctuation appear to remain the most viable explanation for
the knots observed in protostellar outflows.  We do not see any
evidence for disruption of the jet by pinch modes.
 
In the case of pulsed cooling jets, we find the primary effect of a
toroidal magnetic field is to confine shocked jet material to the
axis, preventing it from begin ejected into the cocoon, and leading to
higher postshock densities in the pulses in comparison to purely
hydrodynamic models.  This result is in contrast to the expectation of
planar radiative shock models, in which the addition of a magnetic
field leads to {\em lower} postshock densities.  Our results indicate
it is important to account for multidimensional effects in the study of
magnetized cooling jets.  Toroidal confinement also leads to radial
variation of quantities in the pulses, even if the density and velocity
initially are constant with radius.  The radial variation of, e.g. the
density, in the pulses is large enough (a factor of 80 in a strongly
magnetized jet) that it will likely affect the computation of the
resulting emission properties.

While there are many uncertainties in the magnetic field strength and
topology implied by the observations,  our present results make it
clear that MHD models of protostellar jets need to be seriously
investigated.  This is particularly true since we find that average
values of the magnetic field which correspond to a plasma $\beta$ much
larger than one (and which therefore one would infer to be too weak
to be important) can still make a significant difference in the dynamics
and physical conditions associated with the jet, and therefore
its emission properties as well.
It is thus important to
understand the asymptotic structure of the field produced by the
mechanism which drives the outflow.

\vspace{0.5cm}

We thank E. de Gouveia Dal Pino for comments on an earlier
version of the manuscript, and an anonymous referee for
suggested improvements.  JS acknowledges support from the DOE
through grant DFG0398DP00215.
PH acknowledges support from the National Science Foundation
through grant AST-9318397 and AST-9802955 to the University of
Alabama.

\newpage

\section{Figure Captions}

\figcaption{
Radial profile of the toroidal magnetic field (solid line)
and thermal pressure (dashed line) in the jet beam.}

\figcaption{
Images of the logarithm of the density at the final time in adiabatic
overdense jets with no magnetic field -- $\beta_\infty ^{Ad}$ -- (top
panel), and weak -- $\beta_1 ^{Ad}$ -- (middle panel) or strong
--$\beta_{1/4}^{Ad}$ -- (bottom panel) toroidal field.  A linear color
scale between $\log n = 0.97$ and 4.32 is used, where $n$ is the
particle number density in cm$^{-3}$.  The maximum density in each case
is $\log n = 4.05$, 4.32 and 3.48, respectively.  The initial density
of the jet and ambient medium are $\log n_j = 3$ and $\log n_a = 2$.
The axes are labelled in units of the initial jet radius; note only the
last $50R_j$ of the computation is shown in each case.  For typical
protostellar jet parameters, $\tau_0 = 152$~yrs and $100R_j =
0.08$~pc.}

\figcaption{
Images of the logarithm of $B_{\phi}$ for the
magnetized jets shown in Figure 2.  A linear color map between $\log
(B_{\phi }/\sqrt{4 \pi}) = -2$ and 1.1 is used.  The maximum values are
$\log (B_{\phi}/\sqrt{4 \pi}) = 0.54$ and 1.1, respectively.}

\figcaption{
Images of the logarithm of the density at the final time in cooling
overdense jets with no magnetic field -- $\beta_\infty ^{Cl}$ -- (top
panel), and weak -- $\beta_1^{Cl}$ -- (middle panel) or strong --
$\beta_{1/4}^{Cl}$ -- (bottom panel) toroidal field.  A linear color
scale between $\log n = 0.30$ and 5.6 is used.  The maximum density in
each case is $\log n = 4.2$, 5.6 and 5.4, respectively.}

\figcaption{
Images of the logarithm of $B_{\phi}$ for the magnetized
jets shown in Figure 4.   A linear color map between $\log (B_{\phi
}/\sqrt{4 \pi}) = -2$ and 1.4 is used.  The maximum values are $\log
(B_{\phi}/\sqrt{4 \pi}) = 1.16$ and 1.4, respectively.}

\figcaption{
Radial profiles of density, axial velocity, temperature, and
toroidal magnetic field in cooling steady jet simulations at $z \sim
68.5R_j$ from $\beta^{Cl}_{\infty}$ (solid), $\beta^{Cl}_{1}$ (dotted),
$\beta^{Cl}_{1/4}$ (long dash), $\beta^{Cl}_{ff}$ (dot long dash),
$\beta^{Cl}_{Rm}$ (dot short dash), and $\beta^{Cl}_{\phi z}$ (short
dash). The density, axial velocity, and temperature are scaled relative
to the ambient density, $\rho_a$, the fiducial speed, $a_0$, and
temperature $T_0=10,000$~K.}

\figcaption{
Time history of the logarithm of the density during the
propagation of an overdense, cooling, strongly magnetized jet.  A
linear color scale between $\log n = -0.06$ and 5.4 is
used.}

\figcaption{
Logarithm of the temperature (top panel), velocity divergence
(middle panel), and velocity vectors (bottom panel) near the head of the
overdense, cooling, strongly magnetized $\beta_{1/4}^{Cl}$ jet, at time
$1.8\tau_0 \approx 274$~yrs.  A
linear color map is used for the temperature between $\log T = 1.87$
and 5.8.  Only negative velocity divergence values are shown, sharp
filaments indicate the location of shocks in the flow.  The maximum
length of velocity vectors is scaled to 350~km~s$^{-1}$.}

\figcaption{
Images of the logarithm of the density at the final time in cooling
overdense jets.  Top
panel: a strong toroidal field -- $\beta_{1/4}^{Cl}$ -- which peaks at
$r=0.9R_{j}$.  Second panel: a  strong toroidal field --
$\beta_{Rm}^{Cl}$ -- which peaks at $r=0.2R_{j}$.  Third panel: a
strong toroidal field -- $\beta_{\phi z}^{Cl}$ -- which peaks at
$r=0.9R_{j}$ combined with a weak axial field.  Bottom panel:  a
force-free toroidal field -- $\beta_{ff}^{Cl}$.  A linear color scale
between $\log n = 0.36$ and 5.4 is used.  The maximum density in each
case is $\log n = 5.36$, 4.93, 4.33 and 5.36, respectively.  From top to
bottom the jets are shown at times of $1.8\tau_0$, $1.8\tau_0$, $1.7\tau_0$,
and $1.7\tau_0$ respectively}

\figcaption{
Images of the logarithm of $B_{\phi}$ for the magnetized
jets shown in Figure 9.  A linear color map between $\log
(B_{\phi}/\sqrt{4 \pi}) = -2$ and 1.5 is used.  The maximum values are
$\log (B_{\phi}/\sqrt{4 \pi}) = 1.37$, 1.22, 1.22 and 1.49,
respectively.}

\figcaption{
Images of the logarithm of the density at the final time for the pulsed
jets: $\beta_{\infty}^{P}$ -- no magnetic field -- (top panel), $\beta_{1}^{P}$
-- equipartition toroidal field -- (middle panel), and $\beta_{1/4}^{P}$ --
strong toroidal field --  (bottom panel).  A linear color scale between
$\log n = -1.4$ and 5.3 is used, where $n$ is the number density in
cm$^{-3}$.  The maximum density in each case is $\log n = 4.46$, 4.65,
and 5.29 respectively.  The initial density of the jet and ambient
medium is $\log n_{j} = 3$ and $\log n_{a} = 2$ respectively.  The axes
are labelled in units of the initial jet radius; note the vertical
scale is expanded by a factor of two for clarity.}

\figcaption{
Images of the logarithm of $B_{\phi}$ for the magnetized pulsed jets
$\beta_{1}^{P}$ and $\beta_{1/4}^{P}$ shown in Figure 11.  A linear color map
between $\log (B_{\phi }/\sqrt{4 \pi}) = -2$ and 1.3 is used.  The
maximum values are $\log (B_{\phi}/\sqrt{4 \pi}) = 1.2$ and 1.3,
respectively.}

\figcaption{
Axial profiles along the axis ($r=0$) of density, pressure, velocity,
and magnetic field for pulsed jets $\beta_{\infty}^{P}$ (solid line),
$\beta_{1}^{P}$ (dotted line) and $\beta_{1/4}^{P}$ (dashed line) at $t =
1.2\tau_0$.  Only the region between $20R_{j}$ and $45R_{j}$ from the
jet nozzle is shown.}

\figcaption{
Images of the logarithm of the density at the final time for the pulsed
jets: $\beta_{1/4}^{P}$ -- strong toroidal field peaked at $r_m=0.9R_j$
-- (top panel), $\beta_{Rm}^{P}$ -- equipartition toroidal field peaked at
$r_m=0.2R_j$ -- (middle panel) and $\beta_{\phi z}^{P}$ -- equipartition
toroidal field peaked at $r_m=0.8R_j$ and a uniform axial field --
(bottom panel).  A linear color scale between $\log n = -2.1$ and 4.9
is used.  The maximum density in each case is $\log n = 4.65$, 4.86,
and 4.35 respectively.  Note the vertical scale is expanded by a factor
of two for clarity.  From top to bottom the jets are shown at times
of 1.6, 1.6, and $1.7\tau_0$ respectively.}

\figcaption{
Images of the logarithm of $B_{\phi}$ for the magnetized pulsed jets
$\beta_{1/4}^{P}$, $\beta_{Rm}^{P}$, and $\beta_{\phi z}^{P}$ shown
in Figure 14.
A linear color map between $\log (B_{\phi }/\sqrt{4 \pi}) = -2$ and 1.3
is used.  The maximum values are $\log (B_{\phi}/\sqrt{4 \pi}) = 1.2$,
1.3, and 1.1, respectively.}

\figcaption{
Radial profiles of density, radial velocity, thermal pressure, and
toroidal plasma $\beta$ in pulses at $z \sim 50R_j$ from  $\beta_{\infty}^{P}$
(solid line), $\beta_{1}^{P}$ (dotted line), $\beta_{Rm}^{P}$ (dashed-dotted
line), $\beta_{\phi z}^{P}$ (short dashed line), and $\beta_{1/4}^{P}$ (long
dashed line).  The density, radial velocity, and thermal pressure are
scaled relative to the ambient density, $\rho_a$, the fiducial speed
$a_0$, and the inlet thermal pressure at the jet edge, $p_0$.}

\newpage
\begin{center}
{\bf TABLE 1} \\ 
\vspace{0.1cm}
 Simulation Parameters For Steady Jets \\
\vspace{0.1cm}
\begin{tabular}{cccccccc} \hline \hline
Simulation &
$M_{j,ms}^{~(a)}$ &
$\bar{M}_{j,ms}^{~(b)}$ &
$M_{a,ms}$ &
$P_{j}^{~(a)}/P_{0}$ &
$r_m/R_{j}$ &
$\beta^{~(c)}_{\phi,m}$ &
$\beta_z^{~(c)}$  \\ \hline
$\beta^{Ad}_{\infty}$   & 63.2 & 63.2 & 20   & 1    & ---  &  ---  & ---  \\ 
$\beta^{Ad}_{1}$        & 43.9 & 48.3 & 20   & 2    & 0.9  & 1     & ---  \\
$\beta^{Ad}_{1/4}$       & 22.4 & 25.5 & 20   & 8    & 0.9  & 0.25  & ---  \\
$\beta^{Cl}_{\infty}$   & 63.2 & 63.2 & 283  & 1    & ---  &  ---  & ---  \\
$\beta^{Cl}_{1}$        & 43.9 & 48.3 & 283  & 2    & 0.9  & 1     & ---  \\
$\beta^{Cl}_{1/4}$       & 22.4 & 25.5 & 283  & 8    & 0.9  & 0.25  & ---  \\
$\beta^{Cl}_{ff}$       & 14.1 & 30.1 & 283  & 1    & 0.05 & 0.062 & ---  \\
$\beta^{Cl}_{Rm}$       & 19.2 & 42.0 & 283  & 11   & 0.2  & 0.25  & ---  \\
$~~{\beta^{Cl}_{\phi z}}^{~(d)}$ & 20.9 & 24.5 & 18.3 & 8 & 0.8  & 0.25  & 1    \\ \hline
\end{tabular}
\end{center}
$^{a}$ on jet axis in first computational zone \\
$^{b}$ linear average \\
$^{c}$ $\beta=1$ coresponds to $B = 58.8 \mu G$ \\
$^{d}$ 10 zones/R$_j$ \\

\newpage
\begin{center}
{\bf TABLE 2} \\
\vspace{0.1cm}
 Simulation Parameters For Pulsed Jets \\
\vspace{0.1cm}
\begin{tabular}{cccccccccc} \hline \hline
Simulation &
$M_{j,ms}^{~(a)}$ &
$\langle{M}_{j,ms}\rangle^{(b)}$ &
$\langle{M}_{j,s}\rangle^{(b)}$ &
$M_{a,ms}$ &
$P_{j}^{~(a)}/P_{0}$ &
$r_m/R_{j}$ &
$\beta^{~(c)}_{\phi,m}$ &
$\beta_z^{~(c)}$ &
$\langle\beta_{\phi}\rangle^{(b)}$  \\ \hline
$~\beta_{\infty}^{P}$ & 63.2 & 63.2 & 63.2 & 283  & 1   & --- &  --- & --- & $\infty$ \\
$\beta_{1}^{P}$       & 43.9 & 50.5 & 68.9 & 283  & 2   & 0.9 & 1    & --- & 6.7 \\
$\beta_{1/4}^{P}$      & 22.4 & 27.3 & 46.4 & 283  & 8   & 0.9 & 0.25 & --- & 8.2 \\
$~\beta_{Rm}^{P}$     & 33.6 & 57.2 & 60.7 & 283  & 3.5 & 0.2 & 1    & --- & 16.4 \\$\beta_{\phi z}^{P}$  & 34.5 & 38.3 & 64.6 & 18.3 & 2.2 & 0.8 & 1    & 1   & 10.2\\
\hline
\end{tabular}
\end{center}
$^{a}$ on jet axis in first computational zone \\
$^{b}$ area weighted average \\
$^{c}$ $\beta=1$ coresponds to $B = 58.8 \mu G$ \\

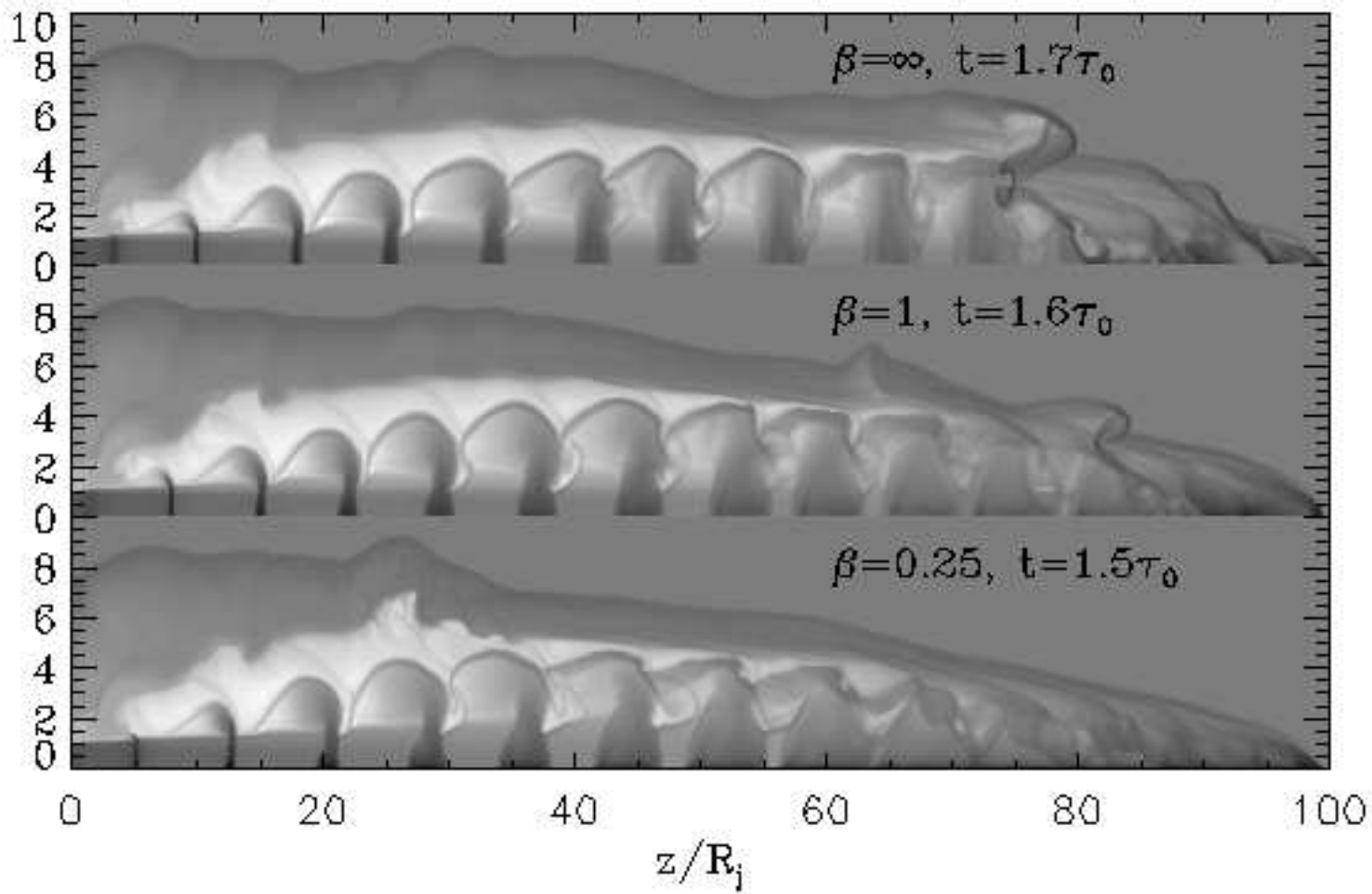
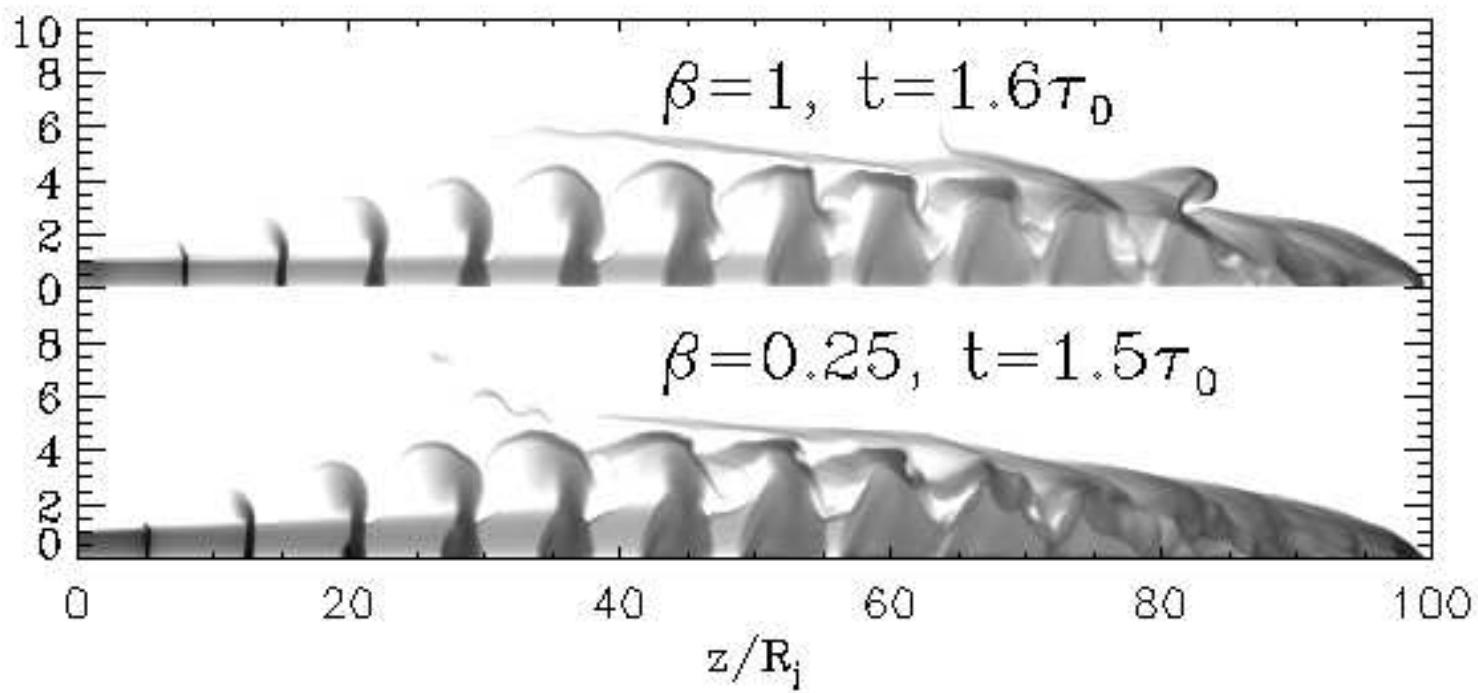
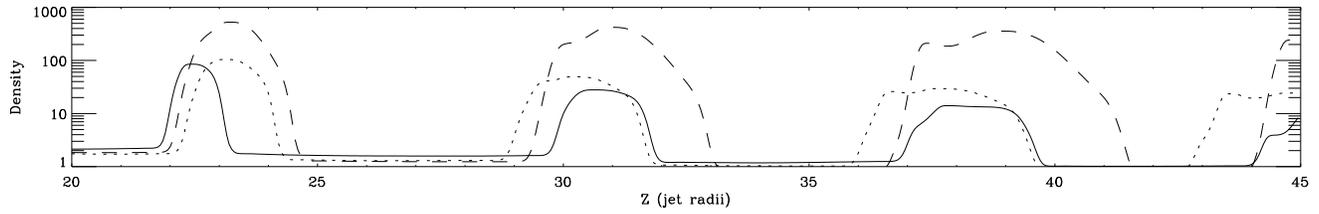
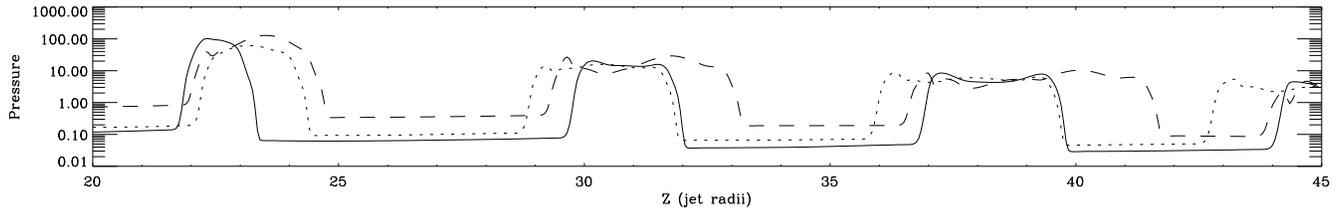
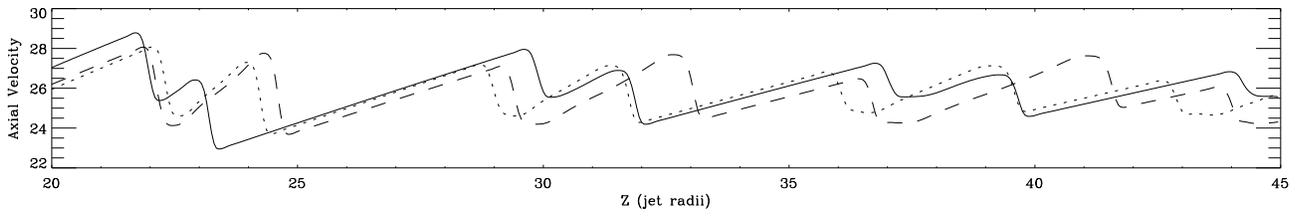
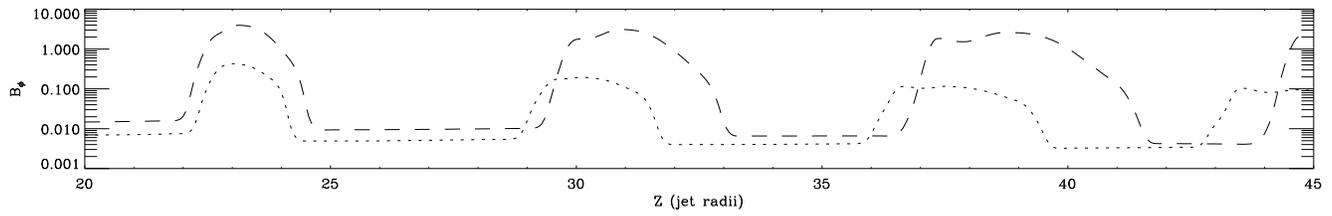
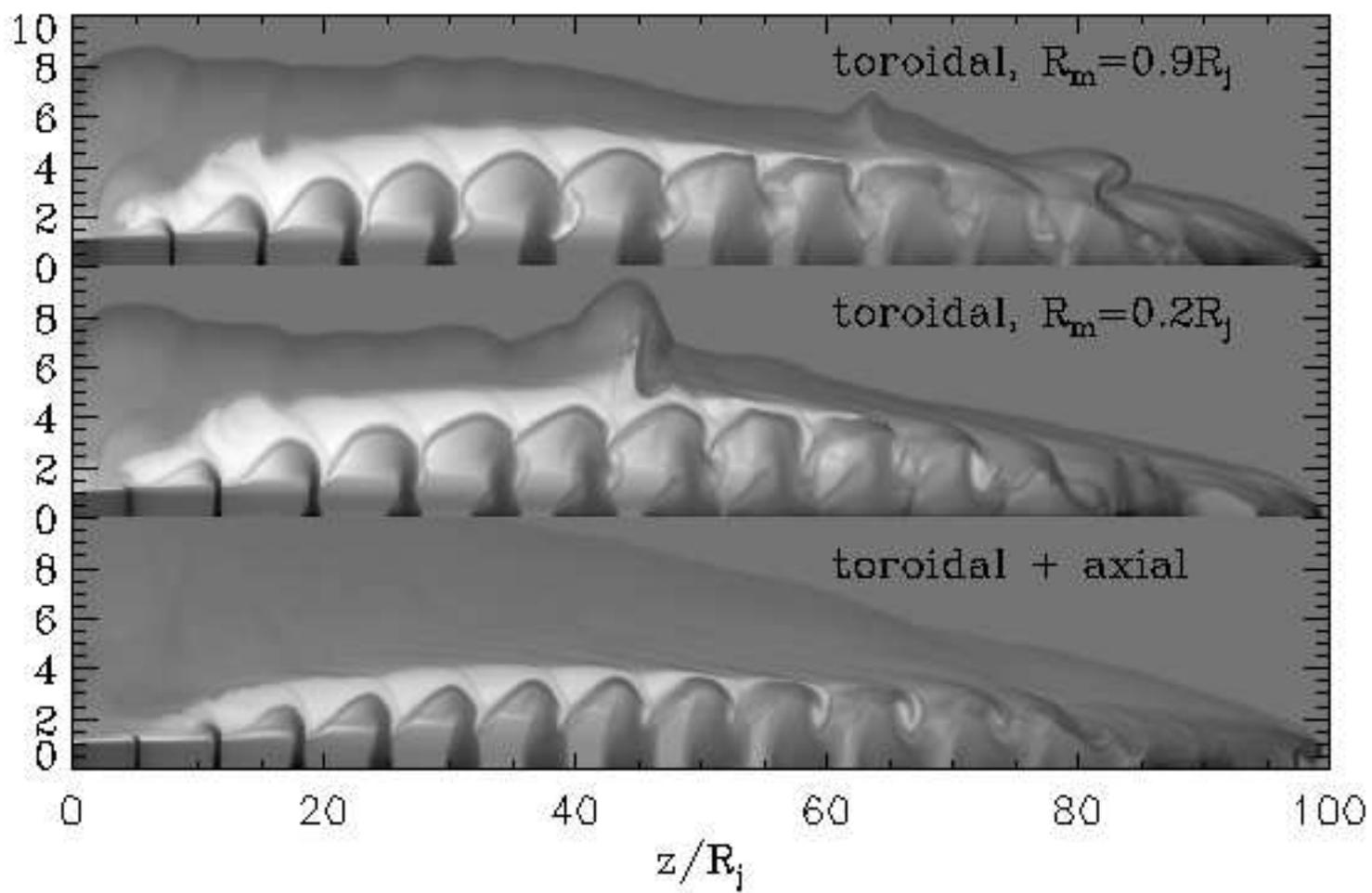
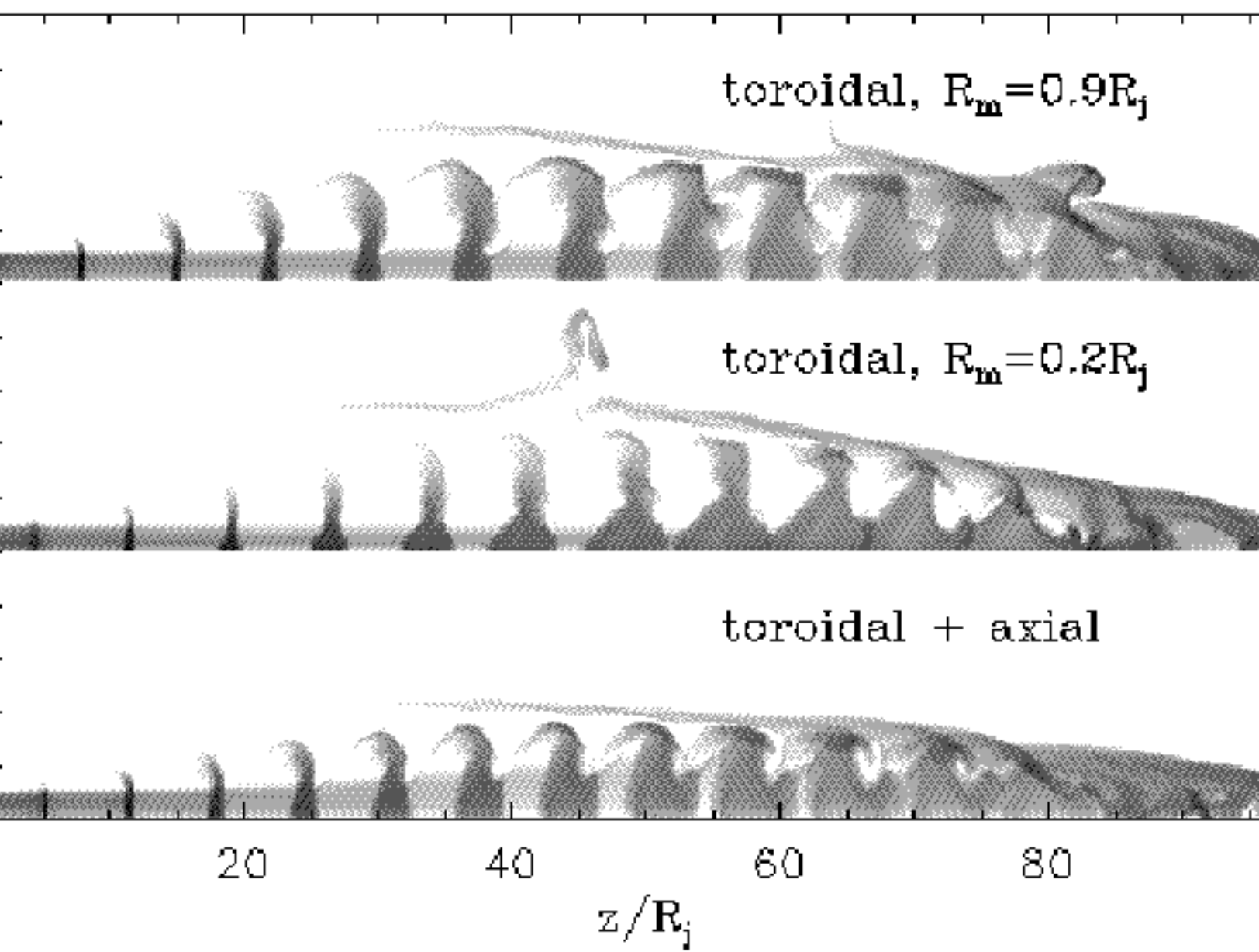
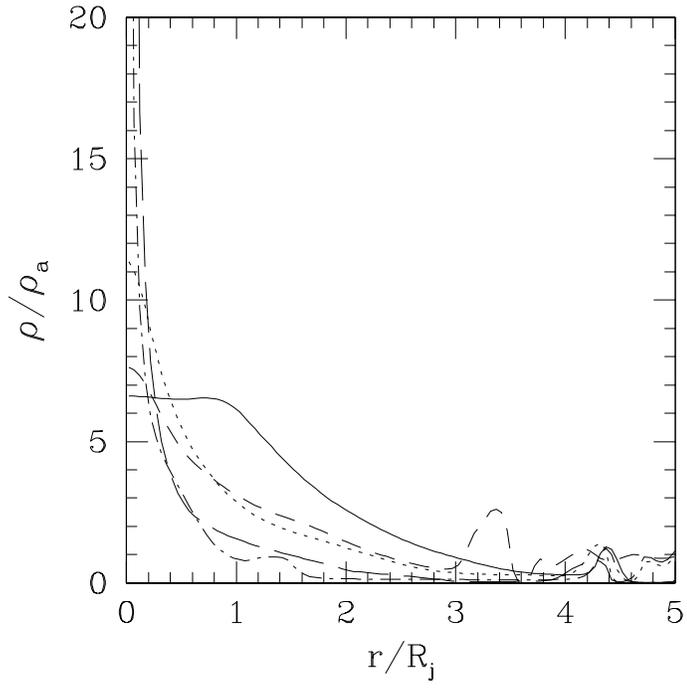
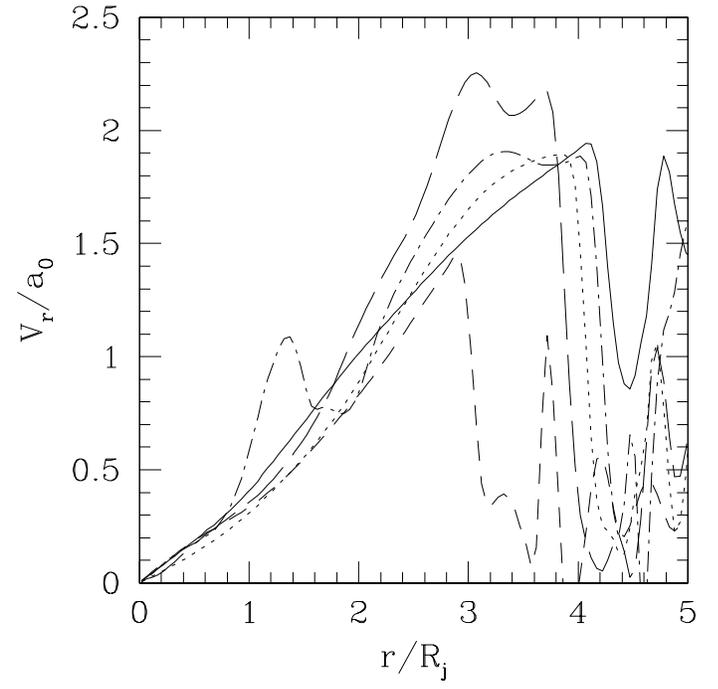
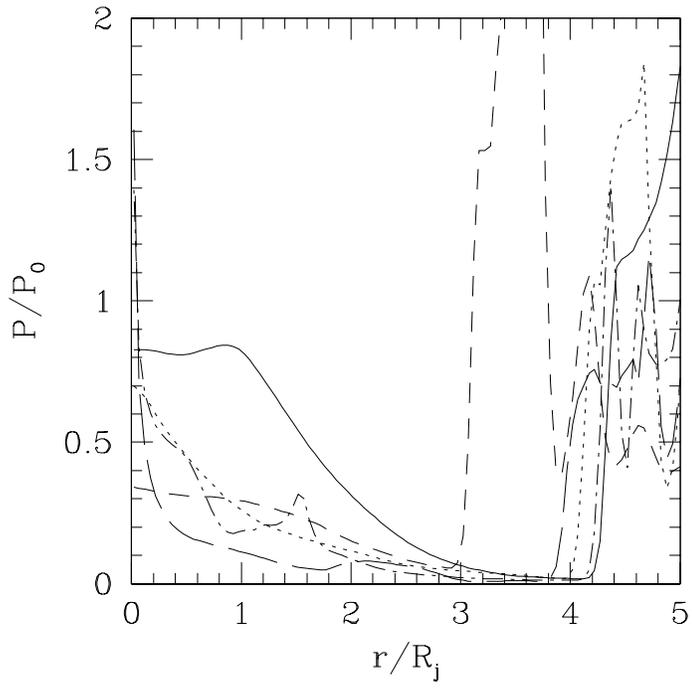
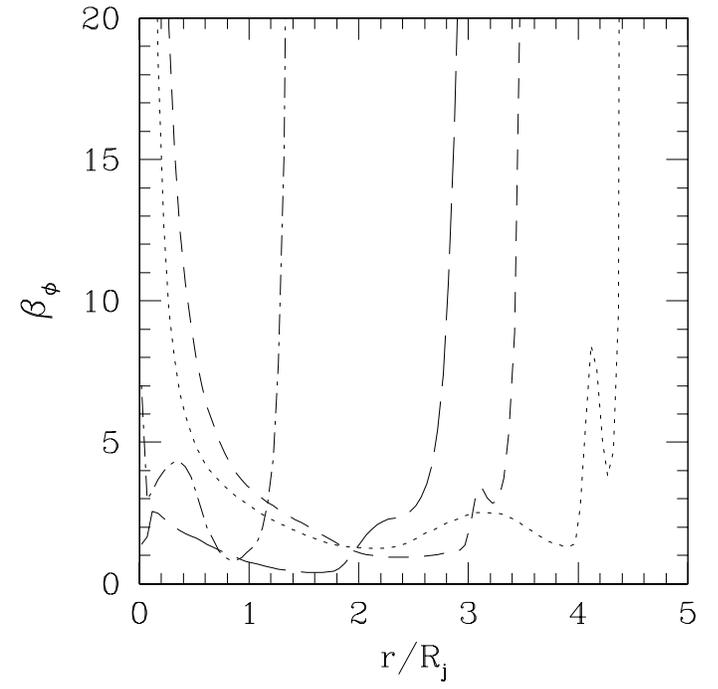
\end{document}